\documentclass[12pt,preprint]{aastex}

\usepackage{natbib}  \newcommand{\um}{$\micron$}
 
\newcommand{\msun}{M$_\odot$} \newcommand{\sfr}{M$_\odot$\,yr$^{-1}$}

\shorttitle{Dust-Corrected Star Formation Rates of Galaxies. II} 
\shortauthors{Hao et al.}

\begin{document}


\title{Dust-Corrected Star Formation Rates of Galaxies. II.  Combinations
of Ultraviolet and Infrared Tracers}


\author{Cai-Na Hao\altaffilmark{1}, Robert C. Kennicutt, Jr.\altaffilmark{2,3}, 
Benjamin D. Johnson\altaffilmark{4,2}, Daniela Calzetti\altaffilmark{5}, Daniel A. Dale\altaffilmark{6},
John Moustakas\altaffilmark{7}}


\altaffiltext{1}{Tianjin Astrophysics Center, Tianjin Normal University, Tianjin
300387, China}
\altaffiltext{2}{Institute of Astronomy, University of Cambridge, Madingley
Road, Cambridge CB3 0HA, UK} 
\altaffiltext{3}{Steward Observatory, University of Arizona, Tucson, AZ 85721, USA}
\altaffiltext{4}{Institute d'Astrophysique de Paris, 98bis Bd Arago, Paris 75014, France}
\altaffiltext{5}{Department of Astronomy, University of Massachusetts,
Amherst, MA 01003, USA} 
\altaffiltext{6}{Department of Physics, University of Wyoming, Laramie, WY 82071, USA}
\altaffiltext{7}{Center for Astrophysics and Space Sciences, University of California,
San Diego, 9500 Gilman Drive, La Jolla, California, 92093-0424, USA}


\begin{abstract} 

We present new calibrations of far-ultraviolet (FUV) attenuation as derived
from the total infrared to FUV luminosity ratio (IRX) and the FUV-NUV color. We
find that the IRX-corrected FUV luminosities are tightly and linearly
correlated with the attenuation-corrected H$\alpha$ luminosities (as measured
from the Balmer decrement), with a rms scatter of $\pm 0.09$ dex. The ratios of
these attenuation-corrected FUV to H$\alpha$ luminosities are consistent with
evolutionary synthesis model predictions, assuming a constant star formation
rate over 100 Myr, solar metallicity and either a Salpeter or a Kroupa IMF with
lower and upper mass limits of 0.1 and 100\msun.  The IRX-corrected FUV to
Balmer-corrected H$\alpha$ luminosity ratios do not show any trend with other
galactic properties over the ranges covered by our sample objects.  In
contrast, FUV attenuation derived from the FUV-NUV color (UV spectral slope)
show much larger random and systematic uncertainties. When compared to either
Balmer-corrected H$\alpha$ luminosities or IRX-corrected FUV luminosities the
color-corrected FUV luminosities show $\sim 2.5$ times larger rms scatter, and
systematic nonlinear deviations as functions of luminosity and other
parameters.  Linear combinations of 25\um\ and 1.4GHz radio continuum
luminosities with the observed FUV luminosities are also well correlated with
the Balmer-corrected H$\alpha$ luminosities. These results provide useful
prescriptions for deriving attenuation-corrected star formation rates of
galaxies based on linear combinations of UV and IR or radio luminosities, which
are presented in convenient tabular form.  Comparisons of our calibrations
with attenuation corrections in the literature and with dust attenuation
laws are also made.

\end{abstract}




\section{INTRODUCTION \label{sec:intro}}

The ultraviolet (UV) continuum of galaxies traces young massive stars,
and hence serves as one of the most commonly used star formation rate (SFR)
indicators (e.g., Kennicutt 1998, hereafter K98). However, the severe dust
attenuation in the UV significantly weakens its power for measuring the SFR.
In nearby star-forming galaxies, the dust attenuation in the UV
can vary from zero to several magnitudes (e.g., Buat et al. 2005; Burgarella et
al.  2005). For near-UV (NUV) selected and far-infrared (FIR) selected
samples, Buat et al. (2005) found that the median dust attenuations in the far-UV
(FUV) are 1.1 mag and 2.9 mag respectively (see also Burgarella et al.  2005),
corresponding to 64\% and 93\% obscuration of the FUV light respectively.
Corrections for dust attenuation therefore are essential before the UV
emission is used to probe the SFR. 

Commonly two approaches are used to estimate the dust attenuation in the UV.
The power-law slope of the UV continuum $\beta$ ($f_\lambda \propto \lambda^\beta$,  
at wavelengths 1300-2600$\AA$; Calzetti et al.  (1994)) was
the first proposed UV dust attenuation indicator (e.g., Kinney et al. 1993;
Calzetti et al. 1994; Meurer et al. 1999), and is based on the assumption that
the intrinsic slope $\beta$ is only sensitive to the recent star formation
history and any deviation from the intrinsic value is caused by dust
attenuation (e.g., Calzetti et al.  1994; Meurer et al. 1995). However, once 
similar analysis is extended from starburst galaxies to normal star-forming
galaxies, $\beta$ will depend on the mean age of the dust-heating population,
the dust/star geometry, and the dust properties (e.g., Witt, Thronson \& Capuano 1992;
Witt \& Gordon 2000; Granato et al. 2000; Burgarella et al. 2005).

The other UV dust attenuation estimator is the ratio of IR to UV emission.
This method is built on an energetic budget consideration (e.g., Meurer et al.
1999; Buat et al. 1999).  According to the energy balance argument, all the
starlight absorbed at UV and optical wavebands by the interstellar dust is
re-emitted in the IR, so the combination of the IR luminosity and the observed
UV luminosity should be able to probe the dust-free UV luminosity (e.g., Wang
\& Heckman 1996; Heckman et al. 1998), which then can be used in turn to
estimate the dust attenuation by comparing to the observed UV luminosity.
Compared to $\beta$, the IR/UV ratio is a more reliable dust attenuation
indicator because it is almost independent of the dust properties and the
relative distribution of dust and stars (e.g., Buat \& Xu 1996; Gordon et al.
2000; Witt \& Gordon 2000; Cortese et al. 2008). However, the IR/UV ratio does
depend on the age of the dust heating populations (e.g., Cortese et al. 2008;
B. D. Johnson et al. 2011, in preparation). The non-ionizing UV emission is
mainly from stars younger than a few hundred million years, whereas by contrast
the IR emission can be from dust heated by stars over a wide age range,
including not only young stars that emit a bulk of their bolometric
luminosities in the UV but also evolved stars that radiate only a small
fraction of their bolometric energies in the UV or even do not contribute to
the UV emission at all.

Since the introduction of these two methods many efforts have been made to
calibrate the relations between UV attenuation and the IR/UV ratio or $\beta$
(or equivalently FUV-NUV color) or the IR/UV versus $\beta$ relation (e.g. Meurer
et al. 1999; Buat et al. 1999; Gordon et al. 2000; Bell 2002; Kong et al.
2004; Calzetti et al. 2005; Seibert et al.  2005; Cortese et al. 2006; Boissier
et al. 2007; Gil de Paz et al. 2007; Johnson et al. 2007a,2007b; Panuzzo et al.
2007; Salim et al.  2007,2009; Treyer et al. 2007; Boquien et al. 2009; Buat et
al. 2010; Takeuchi et al. 2010).  Most of the calibrations are either purely
based on theoretical models or based on comparisons of multi-wavelength
photometry with stellar population synthesis models.  For example, Meurer et
al. (1999) obtained a relationship between FIR/FUV and the attenuation at FUV
($A_{\rm FUV}$) using the stellar population synthesis models of Leitherer \&
Heckman (1995) and then they derived an $A_{\rm FUV}$--$\beta$ relation using
this FIR/FUV--based $A_{\rm FUV}$ for starburst galaxies. Buat et al. (2005)
derived $A_{\rm UV}$--IR/UV relations for star-forming galaxies at FUV and NUV
wavebands based purely on stellar population synthesis modelling.  Salim et al.
(2007) estimated the UV attenuation by comparing the observed spectral energy
distributions (SEDs) to modelled SEDs and then based on this UV attenuation
they obtained $A_{\rm FUV}$--$\beta$ relations for star-forming galaxies
selected spectroscopically or using NUV-r color.  Other studies that use
different samples or types of galaxies to derive prescriptions for estimating
UV attenuation either from IR/UV or from $\beta$ (or equivalently FUV-NUV
color) include Burgarella et al.  2005, Seibert et al.  2005 and Cortese et al.
2008. As discussed earlier, the IR/UV ratio and the FUV-NUV color are dependent
on the age of the dust heating populations, so calibrations based on different
samples, regardless of the methods adopted, should account for this age effect.
The considerable inconsistencies between the above prescriptions (see Section
\ref{subsec:comparisons}) imply that the age difference of the stellar
populations of the samples used was largely taken into account, and these
prescriptions have been applied widely in the literature to correct for UV
attenuations (e.g., Bell \& Kennicutt 2001; Iglesias-P\'aramo et al.  2006).
However, the reliability and accuracy of these corrections need to be evaluated
using an independent set of attenuation estimates.  

In the literature an independent attenuation estimate, the attenuation in the
H$\alpha$ line (A$_{\rm H\alpha}$) as derived from the Balmer decrement ratio,
has also been used to estimate the attenuation in the UV.  To achieve this, two
methods have been employed. One is based on the idea that the SFR derived from
attenuation-corrected H$\alpha$ luminosity matches that from
attenuation-corrected UV luminosity (e.g., Buat et al. 2002; Treyer et al.
2007). Using this method, Buat et al.  (2002) explored the possibility of using
the observed H$\alpha$/UV ratio to estimate the UV attenuation. Treyer et al.
(2007) employed this method to estimate the UV attenuation and then built an
A$_{\rm FUV}$ -- FUV-NUV relation. The alternative approach is to obtain a
scale factor between the attenuations in the H$\alpha$ line and in the UV by
assuming a dust attenuation law (e.g., Buat et al.  2002).  Both methods are
strongly model-dependent, however, which adds to systematic uncertainties in
the results. The UV attenuation derived using the first method, i.e., by
matching the UV and H$\alpha$ derived SFRs, depends not only on the
Balmer-based attenuation estimates in the H$\alpha$ line but also on the
conversions of H$\alpha$ and UV luminosities to SFRs. On the other hand,
the second method rests heavily on the form of the assumed dust attenuation law.

In this paper we take advantage of the large sets of multi-wavelength
observations of nearby galaxies, including UV, IR and optical spectra derived
Balmer decrement ratios, to calibrate IR/UV and FUV-NUV as UV attenuation
estimators empirically in a self-consistent way.  The calibrations derived this
way are purely empirical, and depend only weakly on the Balmer-derived A$_{\rm
H\alpha}$.

This paper is the second of a series, in which we aim to use multi-wavelength
measurements of nearby galaxies to derive attenuation corrections. In Kennicutt
et al.  (2009; hereafter Paper I), we used the Spitzer Infrared Nearby Galaxies
Survey (SINGS; Kennicutt et al. 2003) and an integrated spectral survey sample
from Moustakas \& Kennicutt (2006; hereafter MK06) to calibrate the
combinations of the optical emission lines (H$\alpha$ and [OII]$\lambda$3727)
and the IR or radio continuum emission as attenuation-corrected SFR indicators.
We found that the dust attenuation corrected H$\alpha$ luminosities, as derived
from the Balmer decrement ratios, can be closely traced by the combinations of
the observed H$\alpha$ emission-line luminosities with 24\um, total IR
(TIR; the bolometric luminosity over the wavelength range 3-1100\um), and
even 8\um\ IR luminosities, with rms scatters $\sim$ 0.1 dex.

In this paper we derive new calibrations for UV attenuation corrections as
derived from the IR/UV ratio and the FUV-NUV color as described above, and test
the consistency of the SFRs computed from IR/UV or FUV-NUV corrected UV
luminosities to those from Balmer-corrected H$\alpha$ luminosities.  Compared
to FUV, NUV can be contributed by more evolved stars ($\sim 1$ Gyr). Therefore
we mainly focus on FUV in this paper. But we also explore the reliability of
using IR/NUV as a dust attenuation estimator as NUV is sometimes used as a SFR
indicator.  The remainder of this paper is organized as follows.  In Section 2
we describe the sample, and describe how the UV flux densities were obtained.
The calibrations of the relationships between the attenuation in the FUV and
the TIR to FUV ratio, the UV color are shown in Section 3. In Section 4 we
apply the calibrations derived in Section 3 to two samples of nearby galaxies
to test the reliability of using these attenuation-corrected FUV luminosities
as SFR indicators.  Second order effects of these attenuation-corrected FUV
luminosities relative to the Balmer-corrected H$\alpha$ luminosities are also
investigated in Section 4.  At last, we explore the possibility of using the
combinations of FUV with monochromatic MIR luminosities 25\um\ (or 24\um) and
radio continuum to estimate dust attenuations in Section 4. In Section 5, the
reliability of using NUV luminosity as a SFR indicator is discussed.  In
Section 6 the limitations and the range of applicability of the use of the
calibrations are discussed.  Comparisons of our calibrations to those in the
literature, to the attenuation curve of Calzetti et al. 2000 and to results
based on SFR matching method are made in Section 6 as well. Finally we
summarize our results in Section 7. For distance estimates, we assume ${\rm
H}_0=70{\rm km\,s^{-1}\,Mpc^{-1}}$ with local flow corrections.

\section{SAMPLE AND DATA \label{sec:sample}}

Our galaxies are drawn from the integrated spectrophotometric survey by MK06
and the SINGS survey. The survey information and the selection criteria were
described in Paper I.  We briefly list them here for completeness. The MK06
survey observed 417 galaxies, which were selected to cover the full range of
optical spectral characteristics found in the local galaxies, using a long-slit
drift-scanning technique.  The reduced spectrum was integrated over a
rectangular aperture that includes most or all of the emission of the galaxy.
The SINGS survey was a multi-wavelength survey from the FUV to the FIR and it
was composed of 75 nearby galaxies that cover wide ranges in morphological
type, luminosity, star formation rate and dust opacity.  In this paper
series we aim to use attenuation-corrected H$\alpha$ luminosities, as derived
from Balmer decrement ratios, to calibrate combinations of optical emission
lines or UV continuum and the IR or radio continuum as dust attenuation
corrected SFR indicators. Therefore several criteria were employed to ensure the quality
of the data (1) galaxies without detectable H$\alpha$ emission were excluded,
(2) in order to obtain reliable measurements of the Balmer decrement ratios,
galaxies with low signal to noise ratio spectra at H$\beta$ (S/N $\lesssim$ 15)
were removed, (3) galaxies classified as active galactic nuclei (AGNs) or
AGN-starburst composites (c.f.  Moustakas et al. 2006; Kewley et al. 2001;
Kauffmann et al. 2003) were excluded, and (4) galaxies with lower than 3$\sigma$
detections at 25\um\ were removed.  Given that we will calibrate the
attenuation estimates in the UV from the TIR/UV ratio and the FUV-NUV color,
FUV and NUV photometric data are needed.  Some of the galaxies used in Paper I
have not been observed by the Galaxy Evolution Explorer (GALEX; Martin et al.
2005) so they are not included in the following analyses.

Similar to Paper I, IR data and the attenuation corrected H$\alpha$
luminosities from H$\alpha$/H$\beta$ ratios are required.  The compilation of
H$\alpha$ fluxes, [NII]/H$\alpha$, H$\alpha$/H$\beta$ and IR data has been
described in detail in that paper. As stated in Paper I, most of the galaxies
in the MK06 sample have not been observed with Spitzer. To maximize the
uniformity of the data, only the Infrared Astronomical Satellite (IRAS) data
were used to derive the TIR luminosities\footnote{As shown in Paper I, the TIR
luminosities derived using IRAS measurements are systematically lower than
those based on Spitzer MIPS measurements by $24\% \pm 23\%$ on average.}, which
was estimated using a weighted sum of IRAS 25\um\ , 60\um\ and 100\um\ emission
following Dale \& Helou (2002; equation (5) of their paper). We will discuss
possible systematics introduced by using IRAS-based TIR luminosities in Section
\ref{subsec:sys}. We also provide calibrations for the combinations of the
monochromatic MIR (IRAS 25\um\ or Spitzer 24\um) luminosities and the observed
UV luminosities in Section \ref{sec:other}. This is useful when FIR data are
not available.  As shown in Paper I, the IRAS 25\um\ luminosity (${\rm
L}(\lambda)=\nu {\rm L}_\nu$) and the Spitzer/Multiband Imaging Photometer for
Spitzer (MIPS) 24\um\ luminosity can be used interchangeably. Given that
Spitzer/MIPS 24\um\ photometry is more sensitive than IRAS we use those
measurements in preference to IRAS for the SINGS galaxies in Section
\ref{sec:other}.  Spitzer/MIPS 24\um\ measurements were also performed for the
central 20\arcsec\ $\times$ 20\arcsec\ regions of the SINGS galaxies in Paper
I, and they are also used in this paper (Section \ref{sec:other}).

For galaxies selected from the MK06 sample, we identified their counterparts in
the UV from GALEX GR4. GALEX performs imaging observations at FUV and NUV
bands, centered at 1528$\AA$\ and 2271$\AA$\ respectively, with a circular
field of view of $1\fdg2$ diameter (Martin et al. 2005). GR4 includes images
taken from the All-sky Imaging Survey (AIS), Medium Imaging Survey (MIS),
Nearby Galaxies Survey (NGS), Deep Imaging Survey (DIS) and Guest Investigators
Survey (GI).  Since artifacts appear much more frequently at the edge of the
field of view, only galaxies within a radius of $0\fdg55$ from the field
center were selected (Morrissey et al. 2007).  Because of the overlap in
coverage of these surveys, sometimes there were more than one image available.
In such cases, we used the one with the longest exposure time in the FUV.
After this process we were left with 98 galaxies (54 in AIS, 4 in MIS, 22 in
NGS, 2 in DIS and 16 in GI) out of the 113 star-forming galaxies selected from
the MK06 sample. But NGC\,1003, which is in AIS, was excluded later because of
artifact contamination in its FUV image. So the final sample consists of
97 objects, with 53 in the AIS survey.

Intensity (-int) and sky background (-skybg) images at FUV and NUV bands were
retrieved from the Multimission Archive at Space Telescope Science Institute
(MAST) website\footnote{http://galex.stsci.edu/GR4/} (Morrissey et al. 2007).
Sky background was subtracted before performing photometry. In order to obtain the UV
fluxes from the same region as H$\alpha$, the rectangular apertures used in the
integrated spectra observations by MK06 were adopted in the UV
photometry\footnote{ We note that the IR fluxes we used were measured by IRAS
and hence they are not aperture-matched. Because of the poor resolution of IRAS
imaging, we cannot estimate the fraction of IR emission outside of the
rectangular apertures from the IRAS imaging data. Instead we used Spitzer/MIPS
24\um\ images for galaxies in both MK06 and SINGS samples to get a rough
estimate of this fraction. By comparing the rectangular photometry to the
global photometry which was adopted from Dale et al. (2007), we found that at
most 10\% of 24\um\ emission is from outside of the rectangular apertures. This
small fraction will not affect our results. }.  Foreground stars within the
apertures, identified by comparing the NUV and the Digitized Sky Survey (DSS)
optical images, were masked during the photometry.  The FUV and NUV magnitude
zero points listed in Morrissey et al.  (2007) were used to convert intensities
into magnitudes. The high-resolution relative response images (-rrhr) from MAST
were used to estimate the uncertainties in the photometry.  The FUV and NUV
flux densities were corrected for Galactic extinction using the Schlegel et al.
(1998) dust map and the Galactic extinction curve derived by Cardelli et al.
(1989) for a total-to-selective extinction of $R_V=3.1$ (c.f., Gil de Paz et
al. 2007).  Specifically, ${\rm A_{FUV}=7.9E(B-V)}$ and ${\rm
A_{NUV}=8.0E(B-V)}$. The Galactic extinction corrected FUV and NUV flux
densities of the MK06 sample are listed in Table 1.  The absolute calibration
uncertainties in FUV (0.05 mag) and NUV (0.03 mag) are added in quadrature.

\begin{figure} 
\epsscale{} 
\plotone{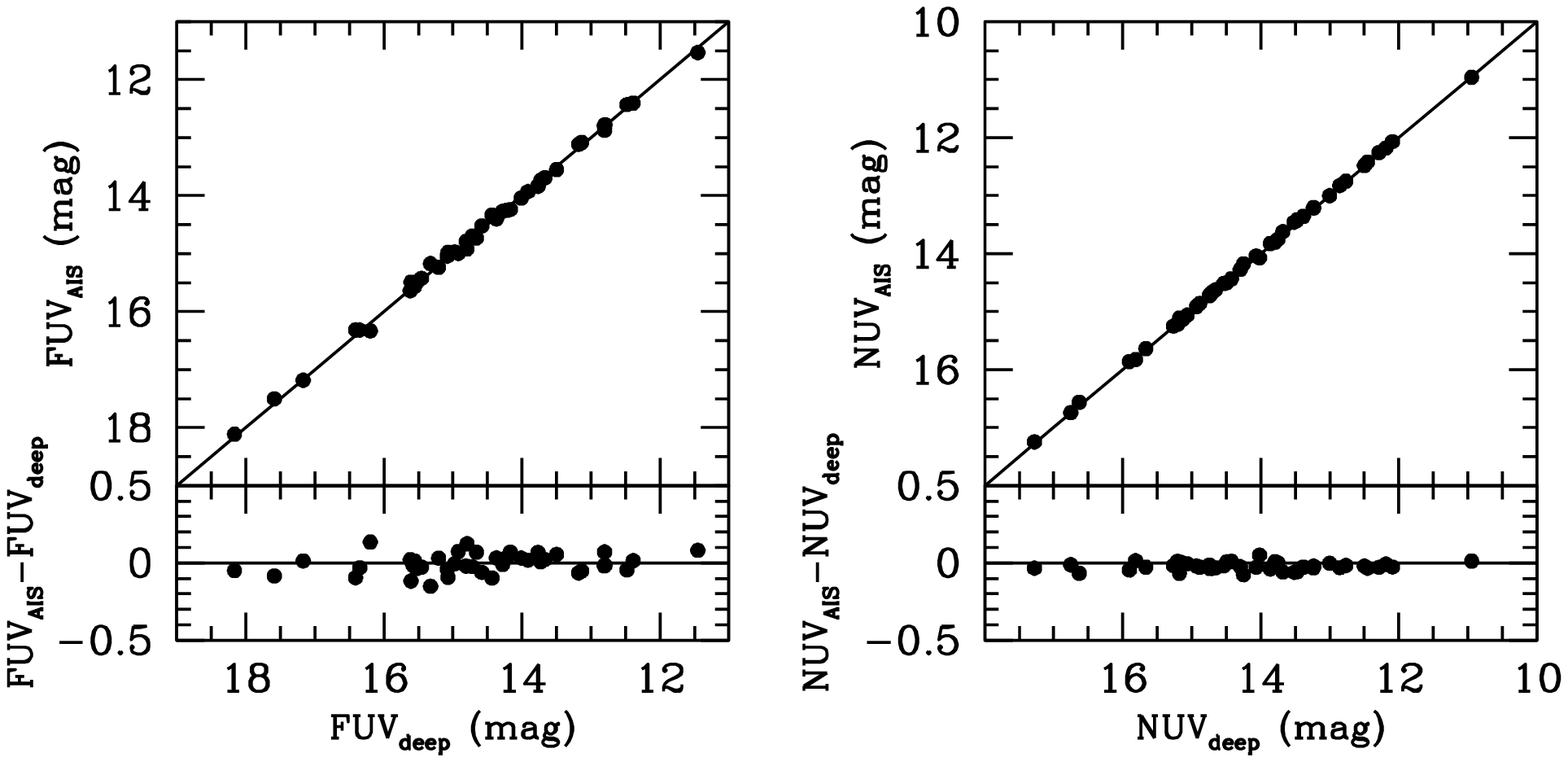}
\caption{Comparison of photometry from GALEX AIS images with that from images with
longer exposure time for MK06 sample at FUV (left) and NUV (right) wavebands.}
\label{comparelongexpAISflux.eps} 
\end{figure}

Since more than half of the MK06 sample galaxies (53 out of 97) were extracted
from the shallow AIS survey,  with exposure times of 100s, 15 times shorter
than the MIS, we tested whether some of the extended UV emission was missed in
AIS images compared to observations with deeper exposures. There are 43 objects
in common between the AIS and the other deeper surveys. We compared their UV
magnitudes measured from the AIS images and those from observations with longer
exposure time in Figure \ref{comparelongexpAISflux.eps}.  It is clear that the
UV measurements based on the AIS images are in excellent agreement, within
$\sim$20\%, with those from deeper exposures, which means that the UV fluxes
measured from the AIS images are not systematically under-estimated within the
adopted rectangular apertures. Consequently they can be used along with the
measures from deeper exposures without introducing any systematic errors.

UV flux densities for the SINGS sample were obtained from Dale et al. (2007).
Those authors calculated the FUV and NUV flux densities from the surface
brightness profiles derived for the GALEX Atlas of Nearby Galaxies (Gil de Paz
et al. 2007). Since Dale et al. (2007) used Li \& Draine (2001) reddening curve
to correct for the Galactic extinction, we redid the Galactic extinction using
the same recipes as those used for the MK06 sample for consistency\footnote{Due
to a mistake in the code, the UV flux densities for the SINGS sample published
in Dale et al. (2007) were virtually foreground corrected according to what Gil
de Paz et al. (2007) used for a reddening curve, same as those used for the
MK06 sample. So they were used directly to compute luminosities and colors.
No corrections were made.}.
Applying the requirement of having both IRAS data and UV flux measurements left
us with 36 objects for the SINGS sample\footnote{Since MIPS 24\um\ measurements
are used in preference to IRAS 25\um\ for the combination of FUV and 25\um\ (or
24\um) for the SINGS galaxies in Section \ref{sec:other}, all star-forming
galaxies in the SINGS sample (see Paper I) with GALEX observations are
included. 47 objects meet this requirement.}. To avoid small number statistics
and the less reliable measurements of H$\alpha$/H$\beta$ ratios for some of the
SINGS objects (see Paper I), we do not use these objects to calibrate the
relations in the following analyses.  However we include them in the comparison
to see how they behave. We also measure the FUV and NUV flux densities for the
central 20\arcsec\ $\times$ 20\arcsec\ regions of the SINGS galaxies. It should
be noted that the large absolute calibration uncertainties (0.15 mag) in Gil de
Paz et al. are included in the quoted errors in FUV and NUV for SINGS sample.

\section{EMPIRICALLY CALIBRATED A$_{\rm FUV}$ -- TIR/FUV, FUV-NUV COLOR RELATIONS \label{sec:calibration}} 

As addressed in Section \ref{sec:intro}, prescriptions of using TIR/FUV as a
dust attenuation indicator are usually derived under various model assumptions.
In this section, we employ a new method to derive such prescriptions
empirically and avoid most of the systematic model dependencies.  Our method is
based on the statistical correlation between the FUV attenuation estimated from
the TIR/FUV ratio and that as derived from the FUV-NUV color.  By deriving the
prescriptions based on a sample of nearby star-forming galaxies, we virtually
assume an average stellar population and dust attenuation curve of a certain
type of galaxies.  Therefore our calibrations are probably not applicable for
other types of galaxies (see Section \ref{sec:uncertainties}).

We adopt the convention of infrared excess IRX from Meurer et al. (1995)
\begin{equation}
{\rm IRX}=\log [{\rm L(TIR)/L(FUV)_{obs}}],
\end{equation}
where the observed FUV luminosity takes the common definition of
monochromatic luminosity ${\rm L(FUV)_{obs}}=\nu {\rm L}_\nu (1528\AA)$.

According to an energy balance argument, the dust attenuation in the FUV, ${\rm
A_{FUV}}$, can be estimated from the IRX using the following form
\begin{equation} 
{\rm A_{FUV}}=2.5 \log (1 + a_{\rm FUV} \cdot 10^{\rm IRX}),
\label{eq:uvextirx} 
\end{equation} 
where $a_{\rm FUV}$ is a scale parameter, whose meaning is explained below.
This approach is similar to that presented in Paper I for the combination of
the observed H$\alpha$ luminosity and the TIR luminosity.  As stated in Paper
I, such an expression is a simple approximation to a much more complicated
radiative transfer process.  To better understand the underlying physical
foundations of eq.  (\ref{eq:uvextirx}), we adopt the relevant equations from
Paper I and adapt them to the case of FUV.

We first define a scale parameter $\eta$, which is
the fraction of the bolometric luminosity that 
is emitted in the FUV, i.e., the inverse of the bolometric correction
\begin{equation}
{\rm L(FUV)_{corr}=\eta_{FUV} L_{bol}},
\label{eq:eta}
\end{equation}
where ${\rm L(FUV)_{corr}}$ represents the intrinsic
(extinction-free) FUV luminosity and ${\rm L_{bol}}$ denotes
the bolometric luminosity.

Under the assumption of a foreground dust screen approximation,
the observed luminosity can be expressed as
\begin{equation}
{\rm L(FUV)_{obs}=L(FUV)_{corr}}\,{\rm e}^{-{\tau_{\rm FUV}}},
\label{eq:tau}
\end{equation}
where $\tau_{\rm FUV}$ is an effective optical depth in the FUV, as defined by
equation (\ref{eq:tau}), and the corresponding
attenuation in magnitudes ${\rm A_{FUV}} = 1.086\,\tau_{\rm FUV}$.

The dust attenuation corrected FUV luminosity is a sum of the 
observed luminosity and the dust-attenuated luminosity
\begin{equation}
{\rm L(FUV)_{corr}= L(FUV)_{obs}+L(FUV)_{corr}}\,(1- {\rm e}^{-{\tau_{\rm FUV}}}).
\label{eq:attenfuv}
\end{equation}

Considering the energy conservation, all the starlight reprocessed by dust
is re-emitted in the IR. So the TIR luminosity can be expressed as
\begin{equation}
{\rm L(TIR) = L_{bol}}\, (1 - {\rm e}^{{-\bar{\tau}}}),
\label{eq:bolometric}
\end{equation}
where $\bar{\tau}$ is the effective opacity for all of the dust-heating starlight,
and is defined as
\begin{equation}
\bar{\tau}= - \ln \frac{\int L_\lambda e^{-\tau_\lambda} d\lambda}
{\int L_\lambda  d\lambda}.
\label{eq:taubar}
\end{equation}
This effective mean opacity depends on the star formation history of the galaxy, 
the dust extinction curve and the relative distribution of stars and dust.

We substitute eqs.\, (\ref{eq:eta}) and (\ref{eq:bolometric}) into eq.\,
(\ref{eq:attenfuv}) to eliminate ${\rm L_{bol}}$ and express ${\rm L(FUV)_{corr}}$ using
the observable luminosities ${\rm L(FUV)_{obs}}$ and L(TIR) and the opacities. After
this exercise, the dust attenuation estimate in the FUV becomes
\begin{equation} 
{\rm A_{FUV}}=2.5 \log [1 + \eta_{\rm FUV}
\cdot 10^{\rm IRX}{{(1 - {\rm e}^{-{\tau_{\rm FUV}}})} \over {(1 - {\rm
e}^{-\overline{\tau}})}}].
\label{eq:uvextbol} 
\end{equation} 

By comparing eq. (\ref{eq:uvextirx}) with eq. (\ref{eq:uvextbol}), it can be
seen that the coefficient $a_{\rm FUV}$ in eq. (\ref{eq:uvextirx}) is the approximation
of the product of the inverse of the bolometric correction $\eta_{\rm FUV}$ and the
factor ${(1 - {\rm e}^{-{\tau_{\rm FUV}}})} \over {(1 - {\rm
e}^{-\bar{\tau}})}$.  

When the FUV-NUV color is used to estimate the FUV attenuation, a linear relation
between ${\rm A_{FUV}}$ and the FUV-NUV color is usually taken
\begin{equation}
{\rm A_{FUV}}=s_{\rm FUV} \cdot {\rm (FUV-NUV)_{obs}} + i_{\rm FUV},
\label{eq:tmp}
\end{equation}
where the slope $s_{\rm FUV}$ and the intercept $i_{\rm FUV}$ are usually determined by 
a linear regression to the observations.
The physical foundations of this relation can be understood as follows. 

The color excess in FUV-NUV, E(FUV-NUV) is defined as
\begin{equation}
{\rm E(FUV-NUV)}={\rm (FUV-NUV)_{obs}-(FUV-NUV)_{int}}={\rm A_{FUV}}-{\rm A_{NUV}},
\end{equation}
where ${\rm (FUV-NUV)_{obs}}$ and ${\rm (FUV-NUV)_{int}}$ denote the observed
and the attenuation-corrected FUV-NUV colors.
E(FUV-NUV) relates to the color excess E(B-V) by the relation
\begin{equation}
{\rm E(B-V) ={ 1 \over (k_{FUV}-k_{NUV})}\, E(FUV-NUV)},
\end{equation}
where ${\rm k_\lambda=A_\lambda/E(B-V)}$ represents an attenuation curve.
So eq. (\ref{eq:tmp}) has the modified form
\begin{equation}
{\rm A_{FUV}={k_{FUV} \over (k_{FUV}-k_{NUV})}} [{\rm (FUV-NUV)_{obs}}-{\rm (FUV-NUV)_{int}}].
\label{eq:modtmp}
\end{equation}
By comparing this modified equation to the original eq. (\ref{eq:tmp}), we can
see that $s_{\rm FUV}={\rm {k_{FUV} \over (k_{FUV}-k_{NUV})}}$, which is the
slope of the UV part of the attenuation curve, and $i_{\rm FUV}/s_{\rm FUV}$ is
equal to the dust-free UV color, ${\rm (FUV-NUV)_{int}}$.

Since the estimated ${\rm A_{FUV}}$ from IRX and FUV-NUV should agree, the predicted
relation between IRX and FUV-NUV color can be obtained by equating eq.
(\ref{eq:uvextirx}) and eq. (\ref{eq:modtmp}). Given that ${\rm
k_{FUV}}$ and ${\rm k_{NUV}}$ cannot be derived separately from the fitting, $s_{\rm
FUV}$ is still used in the IRX -- FUV-NUV relation
\begin{equation}
{\rm IRX}=\log[10^{0.4 s_{\rm FUV} \cdot {\rm [(FUV-NUV)_{obs}-(FUV-NUV)_{int}]}}-1]-\log(a_{\rm FUV}).
\label{eq:irxbeta}
\end{equation}

Before we proceed to derive the parameters $a_{\rm FUV}$, ${\rm
(FUV-NUV)_{int}}$ and $s_{\rm FUV}$ in eq. (\ref{eq:irxbeta}) from the
observations, we discuss the implicit assumptions in our method. The comparison
between eq. (\ref{eq:uvextirx}) and eq. (\ref{eq:uvextbol}) shows that the
coefficient $a_{\rm FUV}$ is the approximation of the product of the inverse of
the bolometric correction $\eta_{\rm FUV}$ and the factor ${(1 - {\rm
e}^{-{\tau_{\rm FUV}}})} \over {(1 - {\rm e}^{-\bar{\tau}})}$.  
By deriving single values for $a_{\rm FUV}$, ${\rm (FUV-NUV)_{int}}$ and
$s_{\rm FUV}$, we virtually study the average properties of this sample,
without taking into account variations in star formation histories and
dust attenuation curves from galaxy to galaxy, which contribute to the
scatters in the observed correlations.

In order to minimize the number of free parameters in the fitting, we first
estimate the intrinsic (i.e., dust-free) FUV-NUV color empirically, which is
achieved by the correlation between the observed FUV-NUV color and the
attenuation in the H$\alpha$ line.  This method is similar to Calzetti et al.
(1994) in which the unattenuated power-law slope of the UV continuum $\beta_0$
was estimated using the correlation between $\beta$ and the Balmer decrement
ratio.  This correlation between the dust reddening in the stellar continuum
and in the gas content (see also Calzetti 1997 and Figure
\ref{uvcolor.Aha.eps} below) implies that the FUV attenuation in the
stellar continuum is proportional to the attenuation in the H$\alpha$ line.
The proportional factor is dependent on the extinction curve and the relative
distribution of stars, gas and dust, which may vary from galaxy to galaxy.
Unfortunately, we cannot study this property on a galaxy-to-galaxy basis based
on current data, so only an average relation is obtained (see Section
\ref{subsec:comdustlaw}).

The FUV-NUV color versus ${\rm A_{H\alpha}}$ relation is shown in Figure
\ref{uvcolor.Aha.eps}.  Throughout this paper the MK06 and SINGS objects are
represented by solid and open circles respectively. From this figure, we can
see that the MK06 and the SINGS samples tell us different stories. The MK06
sample shows that ${\rm A_{H\alpha}}$ is correlated with FUV-NUV color, whereas
the SINGS sample shows almost no correlation between these two quantities after
the two reddest objects are excluded. This is not unexpected given the large
calibration uncertainties of UV photometry and the way we compiled the Balmer
decrement ratios for the SINGS galaxies, as described in Section
\ref{sec:sample} and Paper I.  Therefore only the MK06 sample is used to derive
the dust-free FUV-NUV color, as indicated in Section \ref{sec:sample}.  Since
the scatter is probably intrinsic, which means that the measurement errors are
not the main contributors to the scatter about the regression line, an ordinary
least square bisector fit method is adopted (Isobe et al. 1990; Feigelson
\& Babu 1992). The fitting to the MK06 sample that is plotted as a solid line
in Figure \ref{uvcolor.Aha.eps} gives
\begin{equation} 
{\rm (FUV-NUV)} = (0.539\pm0.029){\rm A_{H\alpha}}+(0.022\pm0.024)
\label{eq:Ahafnc}
\end{equation}
This indicates an intrinsic FUV-NUV color for zero attenuation of
$0.022\pm0.024$ mag. The errors quoted here in the slope and intercept were
yielded by the fitting routine.  The dust-free FUV-NUV color derived here
is in excellent agreement with the value $0.025\pm0.049$ mag derived by Gil de
Paz et al. (2007).  Those authors used the prescriptions given by Buat et al.
(2005) to correct for the dust attenuation in the FUV and NUV using the TIR/FUV
and TIR/NUV ratios, and found that the intrinsic FUV-NUV color for their sample
is $0.025\pm0.049$ mag on average. To understand the average stellar population
of our sample galaxies, we compare the intrinsic FUV-NUV color we derived here with
the stellar population synthesis predictions. By comparing to models
constructed using GALAXEV by Bruzual \& Charlot (2003), we find that our
galaxies are $\sim$ 0.02, 0.03 and 0.04 mag redder in FUV-NUV color than a 10 Gyr old stellar
population with an e-folding timescale of 5 Gyr, 10 Gyr and a constant star
formation history, respectively. For a constant star formation history STARBURST99
(Leitherer et al. 1999; Vazques \& Leitherer 2005) produces consistent result
(see Section \ref{subsec:sys}).

After fixing the term ${\rm (FUV-NUV)_{int}}=0.022$ in eq. (\ref{eq:irxbeta}),
the coefficients $s_{\rm FUV}$ and $a_{\rm FUV}$ can be obtained by fitting eq.
(\ref{eq:irxbeta}) to the data.  Figure \ref{irxfuv.uvcolor.eps} shows the IRX
versus FUV-NUV color relation, with the corresponding $\beta$ labelled on the
top axis.  The relation between FUV-NUV and $\beta$ was derived using the
definition given by Kong et al.  (2004) for our adopted central wavelengths for
FUV and NUV (see Section \ref{sec:sample}), which has the form $\beta=2.32({\rm
FUV-NUV})-2.00$.  In order to have a clearer view of the average properties of
the data, we overplot the IRX median, lower (25\%) and upper (75\%) quartiles in
bins of width 0.2 mag in FUV-NUV color as open squares with error bars in
Figure \ref{irxfuv.uvcolor.eps}.  The solid line represents the chi-square
minimization fit to the MK06 sample with the deviation in the IRX minimized,
which has the form defined by eq.  (\ref{eq:irxbeta}) with $s_{\rm FUV}$ and
$a_{\rm FUV}$ equal to $3.83\pm0.48$ and $0.46\pm0.12$ respectively.
In order to test the effect of the adopted ${\rm (FUV-NUV)_{int}}$ on the
fitting results, we varied ${\rm (FUV-NUV)_{int}}$ by $\pm1\sigma$ about the
best-fitted value, i.e., assumed ${\rm (FUV-NUV)_{int}}=0.046$  and ${\rm
(FUV-NUV)_{int}}=-0.002$.  It turned out that the changes in the best fitted
$s_{\rm FUV}$ and $a_{\rm FUV}$ caused by these new assumptions of the ${\rm
(FUV-NUV)_{int}}$ are within $\pm1\sigma$, and the resulting IRX versus FUV-NUV
color relations cannot be distinguished from that using ${\rm
(FUV-NUV)_{int}}=0.022$ over the ranges of FUV-NUV color and IRX covered by our
sample galaxies (see Section \ref{subsec:limitations}).  Substituting the best
fitted coefficients into eqs.  (\ref{eq:uvextirx}) and (\ref{eq:modtmp}), the
attenuation estimates based on the IRX and the FUV-NUV color become 

\noindent
\begin{equation} 
{\rm A_{FUV}}=2.5 \log [1+ (0.46\pm0.12) \cdot 10^{\rm IRX}], 
\label{eq:irxextfinal} 
\end{equation}
and
\begin{equation} 
{\rm A_{FUV}=(3.83\pm0.48) [{\rm FUV-NUV)_{obs}}-(0.022\pm0.024)}].  
\label{eq:fncextfinal} 
\end{equation}

\section{DUST ATTENUATION CORRECTED FUV LUMINOSITY AS A SFR INDICATOR\label{sec:uvtir}}

In order to evaluate the reliability of the dust attenuation corrected FUV
luminosities derived using the above calibrations as SFR indicators, we compare
the IRX and FUV-NUV color corrected FUV luminosities (according to eq.
(\ref{eq:irxextfinal}) and (\ref{eq:fncextfinal})) with the Balmer decrement
ratio corrected H$\alpha$ luminosities and investigate possible second-order
trends in the residuals as functions of various properties of the galaxies in
this section. Combinations of FUV with monochromatic 25\um\ (or 24\um)
infrared and radio continuum as dust attenuation measures are also explored.
We note that the attenuation calibration methods we employed in Section
\ref{sec:calibration} make only minimal use of the Balmer decrement data. So
when we compare the dust-corrected FUV luminosities based on our calibrations
with the Balmer-corrected H$\alpha$ luminosities, we are not making a circular
argument.

Before we proceed to compare the dust attenuation corrected FUV luminosities
with the H$\alpha$ luminosities, it is interesting to see the behavior of the
uncorrected FUV luminosities as a function of the Balmer decrement corrected
H$\alpha$ luminosities in Figure \ref{lfuvobs.lhacor.eps}, which can help us
understand the importance of the attenuation corrections for this sample. The
straight lines in Figure \ref{lfuvobs.lhacor.eps} are model predictions for
matched SFRs as described below.  Figure \ref{lfuvobs.lhacor.eps} shows that
the uncorrected FUV luminosities can underestimate the SFRs by more than an order
of magnitude, which varies with the intensity of the star formation activity.
Galaxies with different SFRs are not equally attenuated, with
more active galaxies suffering from higher attenuations. This is consistent
with studies in the literature (e.g., Wang \& Heckman 1996; Treyer et al.
2007).  In the following part of this section, the reader will see how much
will be improved with our dust attenuation corrections.

\subsection{Infrared-Corrected FUV Luminosities As SFR Indicators\label{subsec:uvtir}}

We compare the IRX-corrected FUV luminosities with the Balmer decrement ratio
corrected H$\alpha$ luminosities in the left panel of Figure
\ref{comlumiirx.his.eps}.  It can be clearly seen that the IRX-corrected FUV
luminosities correlate tightly and linearly with the Balmer-corrected H$\alpha$
luminosities with a rms scatter of $\pm0.09$ dex.  

In order to quantify the degree of consistency between the SFRs derived from
the attenuation-corrected FUV and H$\alpha$ luminosities, in the left panel of
Figure \ref{comlumiirx.his.eps} we overplot the predicted relations from
different SFR prescriptions. The red short-long dashed line represents the
prediction from the widely used K98's SFR prescriptions.  The calibrations in
K98 assume a Salpeter initial mass function (IMF; Salpeter 1955) with mass
limits 0.1-100${\rm M}_\odot$, and 1990's generation stellar evolution models
(see Kennicutt et al.  1994; Madau et al. 1998).  To compare with the K98's
prediction, we first construct a model under the same assumptions as K98 did,
i.e., Salpeter IMF, constant star formation history lasting for 100 Myr and
solar metallicity (the magenta short dashed line), using Version 5.1 of the
stellar population synthesis model STARBURST99 (Leitherer et al.  1999; Vazquez
\& Leitherer 2005), which uses the state-of-the-art stellar evolutionary
models.

Star-forming galaxies in our sample probably have experienced continuous star
formation for at least 1 Gyr. In order to examine age effects, we build models
with a constant star formation history lasting for 1 Gyr (represented by long
dashed lines).  It is worth noting that for a constant star formation
history the intrinsic FUV to H$\alpha$ luminosity ratio stays constant after 1
Gyr.  Therefore, the model predictions we derive by assuming a constant SFR for
1 Gyr are the same as those under the assumption of a constant SFR for longer
time. Although we assume constant star formation histories here, models with
exponentially declining star formation histories do not give significantly
different results, as long as the SFR does not drop too rapidly.  For example,
for an e-folding timescale of 5 Gyr, the FUV to H$\alpha$ ratio differs from
the constant star formation history by $\leq 4$\% at any given age, according
to our experiments based on stellar population synthesis models constructed
using GALAXEV by Bruzual \& Charlot (2003).

The IMF also affects the FUV/H$\alpha$ ratio. So apart from models with
Salpeter IMF, models with a more realistic Kroupa IMF (Kroupa \& Weidner 2003)
with mass limits of 0.1-100${\rm M}_\odot$ are also explored (represented by
lines in blue).  The FUV and H$\alpha$ luminosities\footnote{In fact we do not
list the luminosities themselves.  Instead we give the scale parameters --
$C_{\rm FUV}$ and $C_{\rm H\alpha}$, which connect luminosities and their
corresponding SFRs by the relation ${\rm SFR(\lambda)}=C_\lambda \cdot {\rm
L(\lambda)}$.} and their ratios are listed in Table 2.  

In the left panel of Figure \ref{comlumiirx.his.eps} the different models
(represented by the colored lines) are difficult to separate, because of the
large dynamic range in luminosities. For that reason we show in the right panel
of Figure \ref{comlumiirx.his.eps} histograms of the attenuation-corrected FUV
to H$\alpha$ luminosity ratio, using a logarithmic scale.  The MK06 and SINGS
samples are represented by the solid and dotted histograms respectively. The
straight lines in color are the same model predictions as those shown in the
left panel of this figure. The median and mean values of the IRX-corrected FUV
to H$\alpha$ luminosity ratios for the MK06 sample are 123.88 and 127.94, which
are in excellent agreement with the model predictions for Kroupa IMF at 100 Myr
and Salpeter IMF at 100 Myr respectively (see Table 2). However, the
FUV/H$\alpha$ ratio only increases by 8\% and 9\% respectively for the same
IMFs but older age ($ \geq 1$ Gyr).

The correlation between the IRX-corrected FUV luminosities and the Balmer-corrected
H$\alpha$ luminosities is so tight and linear that it is useful to test for any second
order trends in the residuals as functions of various galactic properties.
In Figure \ref{fuvirxresi.para1.eps}, we plot the ratios of the IRX-corrected
FUV luminosities to the Balmer-corrected H$\alpha$ luminosities against the
Balmer-corrected H$\alpha$ luminosities (i.e., the dust corrected SFRs; top-left
panel), the Balmer-corrected H$\alpha$ luminosity densities (top-right panel),
the attenuations in the H$\alpha$ line (bottom-left) and the axial ratios b/a
(i.e., inclinations; bottom-right). In Figure \ref{fuvirxresi.para2.eps}, the
optical spectral features -- H$\alpha$ equivalent width EW(H$\alpha$) (top-left
panel), 4000\AA\ break ${\rm D_n}(4000{\rm \AA})$ (top-right panel), gas-phase
oxygen abundance 12+log(O/H) (bottom-left panel) and the dust temperature
indicator -- far-infrared 60\um\ to 100\um\ flux ratio
$f_\nu(60\micron)/f_\nu(100\micron)$ (bottom-right) are explored.  Given that
H$\alpha$ line and continuum emission are SFR and stellar mass tracers
respectively, the integrated H$\alpha$ equivalent width is a measure of
specific SFR (Kennicutt et al. 1994). The ${\rm D_n}(4000{\rm \AA})$ is often
used as a reddening-insensitive rough star formation history indicator of the
stellar populations (e.g., Bruzual 1983; Balogh et al. 1999; Kauffmann et al.
2003) and it correlates with the specific SFR (Brinchmann et al. 2004). In all
panels the dashed line indicates the predicted FUV to H$\alpha$ luminosity
ratio for a Kroupa IMF at age 100 Myr, as shown in Figure
\ref{comlumiirx.his.eps}.

As can be seen from Figure \ref{fuvirxresi.para1.eps} and
\ref{fuvirxresi.para2.eps}, the ratios of the IRX-corrected FUV to the
attenuation-corrected H$\alpha$ luminosities show no or at most marginal trends
with all the parameters investigated.  The rms scatters are about $\pm0.09$
dex, which is even smaller than the average uncertainty in $\log{\rm
[L(FUV)_{IRX,corr}/L(H\alpha)_{corr}]}$ ratio, $\pm0.13$ dex, with the error
introduced by using eq. (\ref{eq:irxextfinal}) to correct for FUV attenuation
included. This demonstrates the reliability of using the combination of the FUV
and TIR luminosities as a SFR indicator. The absence of correlations implies
that over the ranges of the galactic properties probed by our galaxies, either
no trend exists in the dust-free FUV/H$\alpha$ ratios and the IRX method with
the parameters tested above, or any trend in the dust-free FUV/H$\alpha$ ratios
is compensated by that in the IRX method.  Whatever it is, the SFR estimated
from the IRX-corrected FUV luminosity agrees well with that derived from the
Balmer-corrected H$\alpha$ luminosity, and the ratio of these dust-corrected
SFRs is not a function of the galactic properties. Therefore the combined FUV
and TIR luminosities can serve as a very good attenuation-corrected SFR
indicator over the ranges of physical properties covered by our sample objects. 

From Figure \ref{comlumiirx.his.eps} to Figure \ref{fuvirxresi.para2.eps}, we
note that the SINGS galaxies deviate from the MK06 galaxies systematically. The
systematic deviations can be caused either by overestimates of the TIR
luminosities or by underestimates of the attenuation in H$\alpha$ lines.  As we
showed in Paper I (Figure 2 in that paper; see also Dale et al. 2009), for cold
galaxies ($f_\nu(60\micron)/f_\nu(100\micron) \le 0.6 $), the TIR luminosities
measured by IRAS bands are systematically underestimated with respect to the
more sensitive and accurate Spitzer/MIPS bands. Therefore using TIR
luminosities based on Spitzer/MIPS measurements for the SINGS galaxies would
make the SINGS galaxies deviated more. Furthermore, we did not see a
correlation between the deviations and the $f_\nu(60\micron)/f_\nu(100\micron)$
colors in the bottom-right panel of Figure \ref{fuvirxresi.para2.eps}. Another
piece of evidence is that the same deviations show up in the comparison of the
combined FUV+25\um\ luminosities with the Balmer-corrected H$\alpha$
luminosities (Section \ref{sec:other}), and the monochromatic 25\um\ luminosity
measured by IRAS does not suffer from the same problem as the TIR
luminosity\footnote{In paper I, we found that the IRAS 25\um\ luminosities are
in well agreement with the Spitzer/MIPS 24\um\ luminosities.}. Therefore, the
systematic deviations of the SINGS galaxies cannot result from overestimates of the
TIR luminosities by IRAS. Instead, they may be due to underestimates of
the attenuation in H$\alpha$ lines, caused by the noisier H$\beta$ detections.

\subsection{FUV-NUV Corrected FUV Luminosities As SFR Indicators\label{subsec:fnc}}

Similar to the analyses we did for the IRX-corrected FUV luminosities in Section
\ref{subsec:uvtir}, we compare the FUV-NUV color corrected FUV luminosities with
the Balmer decrement ratio corrected H$\alpha$ luminosities in this subsection.
Figure \ref{comlumifnc.his.eps} shows the FUV-NUV corrected FUV luminosities as a
function of the Balmer-corrected H$\alpha$ luminosities (left panel) and the
histogram of the FUV-NUV corrected FUV to the Balmer-corrected H$\alpha$
luminosity ratio (right panel). Compared to the IRX-corrected FUV luminosities
versus the Balmer-corrected H$\alpha$ luminosities relation, the correlation
between the FUV-NUV corrected FUV luminosities and the Balmer-corrected H$\alpha$
luminosities is non-linear and has a much larger rms scatter of $\pm0.23$ dex.

Figure \ref{comlumifnc.his.eps} shows that the relation between the FUV-NUV
color corrected FUV luminosities and the Balmer-corrected H$\alpha$
luminosities is not completely linear. The slope of the relation is less than
unity, which implies that for galaxies forming stars vigorously (i.e., with
high SFRs), the FUV-NUV corrected FUV luminosities will under-estimate the
SFRs. 

To better understand the applicability and limitation of the FUV-NUV corrected
FUV luminosities as SFR measures, apart from the Balmer-corrected H$\alpha$
luminosities, we also examined the ratios of the UV color corrected FUV
luminosities and the Balmer-corrected H$\alpha$ luminosities as functions of
several other galactic properties, as we did for the IRX-corrected FUV
luminosities.  We found that the residuals of the UV color corrected FUV
luminosities relative to the Balmer-corrected H$\alpha$
luminosities\footnote{In order to minimize any systematic effects introduced by
using the Balmer-corrected H$\alpha$ luminosities as references, we also
examined the trends of the residuals of the UV color corrected FUV luminosities
relative to the IRX-corrected FUV luminosities as functions of the above
parameters. We found that these residuals behave similarly to the residuals of
the UV color corrected FUV luminosities with respect to the Balmer-corrected
H$\alpha$ luminosities.} show significant negative correlations with several of
the parameters, including the Balmer-corrected H$\alpha$ luminosity densities,
the attenuations in the H$\alpha$ lines, the FIR color and the EW(H$\alpha$),
as shown in Figure \ref{fuvuvcolorresi.para1.eps} and
\ref{fuvuvcolorresi.para2.eps}.  The negative trends with the FIR color and the
H$\alpha$ equivalent width are similar to what Kong et al. (2004) found.  Those
authors found correlations between the deviation from the average IRX-$\beta$
relation defined by starburst galaxies and the above two parameters, and they
explained these correlations as direct observational evidence for the
dependence of the scatter of the IRX-$\beta$ relation on star formation
history.  Galaxies that are forming stars more actively in the current epoch
than in the past are intrinsically bluer than those with lower current to past
SFR ratios. So the attenuation in those galaxies with warmer FIR color, larger
EW(H$\alpha$) and hence bluer intrinsic FUV-NUV color, would tend to be
under-estimated by the prescription defined by the average population of the
galaxy sample (see eq.  (\ref{eq:fncextfinal})).  Since stars older than $\sim
1$ Gyr do not contribute to UV emission, the above argument is only valid for
galaxies with either continuous star formation history or continuous star
formation history with superposed bursts younger than $\sim 1$ Gyr. For a
continuous star formation history with superposed bursts taking place at more
than $\sim 1$ Gyr ago, the FUV-NUV color behaves as if no bursts occurred at
all in the history.  On the other hand, the correlations of the residuals of
the FUV-NUV corrected FUV luminosities with the H$\alpha$ luminosities, the
H$\alpha$ luminosity densities and the H$\alpha$ attenuations are hard to be
interpreted in terms of the dust-free FUV-NUV color because there is no
evidence for a correlation between the absolute star formation activity and the
intrinsic FUV-NUV color.  The possible reason for these correlations is the
change in the effective attenuation curves (see eq.  (\ref{eq:modtmp})) with
the star formation intensities given that the intensity of the star formation
activity may change the dust properties and the stars/dust geometry, and hence
the effective attenuation curves.  These speculations agree with the argument
by Burgarella et al. (2005) and Boquien et al. (2009), who suggested that both
star formation history and dust attenuation curve effects should be taken into
account when the FUV-NUV color is used as a UV attenuation estimator. In other
words, a simple formalism of using FUV-NUV color as a FUV attenuation measure
(e.g., eq. (\ref{eq:tmp})) is far from enough. Additional constraints on star
formation histories and dust attenuation curves are needed.

\subsection{Other Composite SFR Indicators \label{sec:other}}

Since FIR photometry and hence the TIR luminosity is sometimes not available, it is
worthwhile calibrating the combinations of the FUV and the monochromatic IR
waveband luminosities (e.g., 25\um\ luminosities\footnote{Since the number
of galaxies with available 8\um\ data is small, no attempt was made to
calibrate the combination of UV and 8\um\ luminosity as an attenuation or
a SFR indicator.}) and radio emission at 1.4GHz as attenuation-corrected SFR
measures.

By matching the combination of the observed FUV and the 25\um\ infrared
luminosities with the IRX-corrected FUV luminosities, we obtain a coefficient
of $3.89\pm0.15$ for FUV+25\um\ , which indicates an average L(25\um\ ) to L(TIR)
ratio of $\sim$ 0.12. This is well within the observed range for star-forming galaxies
(c.f., Paper I; Calzetti et al. 2010).
The left panel of Figure
\ref{lumifuv25umradio.lha.eps} shows a tight and linear correlation between the
Balmer-corrected H$\alpha$ luminosities and the combined FUV and 25\um\
 luminosities.  Since the relation between the Balmer-corrected H$\alpha$
luminosity and the IRX-corrected FUV luminosity, which is the reference of the
combined FUV and 25\um\ luminosity, is consistent with the predicted
relation for a constant star formation history lasting for 100 Myr and Kroupa
IMF (see Section \ref{subsec:uvtir}), we plot this model prediction here for
reference (the dashed line). As expected, the data match the model well. The
dispersion is $\pm0.13$ dex.

In this panel, besides SINGS (open circles) and MK06 (filled circles) samples,
we also include the central 20\arcsec\ $\times$ 20\arcsec\ regions of the SINGS
galaxies, as denoted by open squares.  Although the calibration was derived
based on the integrated measurements of MK06 sample, it also compares well to the
centers of the SINGS galaxies. This perhaps is not surprising because of the
large coverage of the centers (typically several square kilo-parsecs).  

To our knowledge, there exists only one such calibration in the literature.
Zhu et al. (2008) calibrated a FUV and 24\um\ combination by comparing to a
FUV-NUV corrected FUV luminosity.  They used the prescription published by
Treyer et al. (2007) to derive the FUV-NUV corrected FUV luminosity and
obtained ${\rm L(FUV)_{corr}=L(FUV)_{obs}+6.31 L(24\micron)}$. The coefficient
of 6.31 is much higher than our value.  In order to understand the discrepancy
between Zhu et al. and our calibration, we compare our eq.
(\ref{eq:fncextfinal}) with the relation given by Treyer et al.  We find that
the relation provided by Treyer et al.  gives systematically lower attenuation
estimates than ours, which implies that a coefficient lower than 3.89 is
required to match the FUV-NUV corrected FUV luminosity derived using Treyer et
al.'s relation. On the contrary, Zhu et al.  obtained a much higher value. We
cannot figure out the reason that leads to a high coefficient obtained by Zhu
et al. (2008).

The combination of the observed FUV and 1.4GHz radio luminosities yields a
similarly tight relation $\pm0.14$ dex as shown in the right panel of Figure
\ref{lumifuv25umradio.lha.eps}, but its correlation with the Balmer-corrected
H$\alpha$ luminosities shows clear nonlinearity, with a slope of $1.09\pm0.02$
from a bisector fit. For the average H$\alpha$ luminosity of our sample, ${\rm
L(H\alpha)_{corr}=10^{41.5}L_\odot}$, the radio-corrected FUV luminosity can deviate
from a linear relation by 0.02 dex.  We conclude that when FIR data are absent,
the monochromatic 25\um\ luminosity and the 1.4GHz radio luminosity can be used
as a slightly less accurate surrogate of the TIR luminosity to correct for the
dust attenuation in FUV. 

\section{CAN NUV LUMINOSITIES BE USED AS A SFR INDICATOR?}

In the previous sections, we have calibrated the FUV luminosity as a SFR
indicator. A relevant question is whether NUV can also serve as a reliable SFR
tracer. In theory, FUV traces emission from stars younger than $\sim 300$ Myr
(e.g., Meurer et al.  2009), whereas NUV can be contributed by stars as old as
$\sim 1$ Gyr. FUV and NUV emissions are both dominated by stars younger than $\sim
100$ Myr though. Our experiment with STARBURST99 shows that for a galaxy with
constant star formation history lasting for more than 1 Gyr, its FUV emission
increases 8\% from 100 Myr to 300 Myr old and only 1\% from 300 Myr to 1 Gyr
old.  The equivalent numbers for NUV are 12\% and 5\%, respectively. We note
that models built from GALAXEV by Bruzual \& Charlot (2003) produce consistent
results.  In other words, the NUV emission of such a galaxy contributed by
stars older than 100 Myr is about 17\% (compared to 9\% for FUV emission),
which makes it a less reliable SFR tracer in galaxies with a significant
fraction of stars being 100 Myr -- 1 Gyr old.  In the past, however, when FUV
observations were rarely available NUV measurements were the only UV SFR
indicators available. Even today, NUV is sometimes used to probe the star
formation activity (e.g., Iglesias-P\'aramo et al. 2006); Recipes for using the
IR/NUV ratio to estimate the attenuation in the NUV have been presented
recently (e.g., Buat et al. 2005; Burgarella et al. 2005; Cortese et al. 2008).
Thus it is useful to assess the reliability of using NUV luminosity as a SFR
indicator, using our new approach and datasets.

Given that NUV emission can be from more evolved stars than FUV, we did not fit
the corresponding coefficient $s_{\rm NUV}$ in eq. (\ref{eq:irxbeta}).  Instead
we estimated $s_{\rm NUV}$ from $s_{\rm FUV}$ according to $s_{\rm NUV}=s_{\rm
FUV}-1$ (i.e.,$s_{\rm NUV}=2.83$)\footnote{In fact when we applied the same
method as in Section \ref{sec:calibration} to obtain $s_{\rm NUV}$ and $a_{\rm
NUV}$, we got a smaller than $s_{\rm FUV}-1$ value but this value is still
consistent with $s_{\rm FUV}-1$ within 1 $\sigma$ error.}. Then $a_{\rm NUV}$
was estimated from fitting eq. (\ref{eq:irxbeta}) with NUV substituted for FUV
data and $s_{\rm NUV}$ fixed. The resulting fitted value of $a_{\rm NUV}$ was
0.27$\pm$0.02. Following the analyses for FUV, we compare the TIR/NUV corrected NUV
luminosities\footnote{When FUV is available, there is no reason to use NUV as a
SFR measure. So the FUV-NUV corrected NUV luminosity is not examined.} with the
Balmer-corrected H$\alpha$ luminosities in Figure
\ref{comluminuvirx.lhacor.rat.his.eps}.  To our pleasant surprise, the combined
NUV and TIR luminosities correlate with the Balmer-corrected H$\alpha$
luminosities nearly as well as FUV, and the scatter in NUV+TIR vs.  L$({\rm
H}\alpha)_{\rm corr}$ is almost the same as that in FUV+TIR vs.  L$({\rm
H}\alpha)_{\rm corr}$ (0.10 vs. 0.09 dex respectively).  The average ratio of
the TIR/NUV-corrected NUV luminosity to the Balmer-corrected H$\alpha$
luminosity is consistent with that expected for a constant star formation
history over $\geq 1$ Gyr for a Kroupa IMF. 

We also examined the trends of the residuals of the combined NUV and TIR
luminosities relative to the Balmer decrement ratio corrected H$\alpha$
luminosities as functions of various parameters.  We find that all the
behaviors shown in NUV are similar to those in FUV.  When 25\um\ infrared and radio
continuum 1.4GHz luminosities are used as substitutes for TIR luminosities, we
obtain coefficients $a_{\rm NUV}$ of 2.26$\pm$0.09 and $41.75\pm2.97 \times 10^{20}$
respectively. The scatters shown in NUV+25\um\ and NUV+1.4GHz versus
L(H$\alpha$)$_{\rm corr}$ are identical to those shown in the same comparisons
for FUV.

From the above analyses we can see that the NUV luminosity can serve as a
rather good SFR indicator for our sample galaxies, and star formation
activities have been going on smoothly in these galaxies over the last 1 Gyr.
However, the use of NUV-based SFRs in galaxies with non-constant star
formation histories during the last 1 Gyr may be problematic.  Our experiment
with the stellar population synthesis modelling code GALAXEV (Bruzual \&
Charlot 2003) shows that for a galaxy with a constant star formation history
lasting for 100 Myr, the NUV emission at 300 Myr is 4.3\% of its value at
100 Myr (compared to 2.6\% for FUV). For a galaxy with high SFR ($\sim 23$
\sfr) lasting for 100 Myr, the NUV emission at 300 Myr cannot be
distinguished from a Milky Way like star-forming galaxy (with SFR $\sim 1$
\sfr).

\section{DISCUSSION\label{sec:uncertainties}}

In previous sections we have derived prescriptions of using TIR/UV and FUV-NUV
as UV dust attenuation indicators by employing a new empirical method.
Specifically, these calibrations were obtained by fitting a TIR/UV versus
FUV-NUV relation, which was defined by assuming that the UV attenuation
estimated from TIR/UV ratio matches that from FUV-NUV color, to a nearby
star-forming galaxy sample. We have demonstrated that the combinations of the
observed FUV and NUV luminosities with the TIR luminosities (i.e., the TIR/UV
corrected UV luminosities) are tightly correlated with the Balmer-corrected
H$\alpha$ luminosities, and hence can serve as good SFR measures. By
referencing to the TIR/UV corrected UV luminosities, we used the 25\um\ (or
24\um) infrared or 1.4 GHz radio continuum luminosities as surrogate of the TIR
luminosity to correct for the dust attenuation in the UV and found that they
are slightly less accurate than the TIR luminosity.  The combinations adopt the
form of ${\rm L(UV)_{corr}=L(UV)_{obs}}+a \cdot {\rm L(\mbox{IR or radio})}$,
and the corresponding attenuation in the UV has the form of ${\rm A_{UV}}=2.5
\log [{1 + {{a \cdot {\rm L(\mbox{\scriptsize IR or radio})}} \over {\rm
L(UV)_{obs}}}}]$, where $a$ is the approximation of the product of the inverse
of the bolometric correction $\eta_{\rm UV}$ and the factor ${(1 - {\rm
e}^{-{\tau_{\rm UV}}})} \over {(1 - {\rm e}^{-\bar{\tau}})}$ (see Section
\ref{sec:calibration}).  The coefficient $a$ and the dispersions of the
combinations relative to the Balmer-corrected H$\alpha$ luminosities are listed
in Table 3.

As we emphasized in Paper I, these composite indices have been calibrated over
limited ranges of galaxy properties and physical environments.  It is therefore
important to understand the range of observations over which these calibrations
are derived, and any systematics which were built into our calibrations.  It is
also useful to compare our results with the calibrations presented in the
literature by different groups. We also reassess the consistency between the
SFR derived from our dust attenuation corrected FUV luminosity and that based
on Balmer-corrected H$\alpha$ luminosity.

\subsection{Range of Applicability\label{subsec:limitations}} 

In Paper I we have already discussed the ranges of the galaxy properties (e.g.,
the attenuation in H$\alpha$, TIR to H$\alpha$ flux ratio, ${\rm D_n}(4000{\rm
\AA})$ and TIR luminosity) covered by our samples. Briefly our sample 
is composed of normal star-forming galaxies in the local universe. Neither
early-type red galaxies nor dusty starburst galaxies are included in this sample,
which is further demonstrated by the UV properties presented in this paper.
The ranges of IRX and FUV-NUV color spanned by our sample galaxies are -0.19 to 2.39 dex 
and 0.07 to 1.06 mag respectively.  This coverage in IRX and FUV-NUV color introduces
uncertainties in the best fitted coefficients, as presented in Section
\ref{sec:calibration}, due to the lack of constraints from objects redder than
1.06 mag and bluer than 0.07 mag.  As a result caution should be exercised when
applying our calibrations to objects outside the above ranges (see Section
\ref{subsec:sys}).

\subsection{Comparisons With Calibrations In The Literature
\label{subsec:comparisons}}

There exist several calibrations for normal star-forming galaxies in the
literature.  These calibrations were obtained under different assumptions about
detailed star formation histories, dust attenuation curves, and the
parameterization of the A$_{\rm FUV}$ -- IRX relation (see below). It is thus
necessary to test the consistency between them.  We compare our calibration for
IRX-based FUV attenuation estimates with those commonly used in the literature
in Figure \ref{compareBuatSBresifuv.eps}.  Here the FUV attenuation (top
panel), the residual of the FUV attenuation estimated using the calibrations
given by other groups relative to that using our calibration (middle panel) and
the normalized (percentage) residual (bottom panel) are plotted as a function
of IRX.  The black solid line denotes our calibration.  The calibrations
presented by other authors are shown with different line types as denoted in
the top panel.  Specifically, the commonly used calibrations for star-forming
galaxies by Buat et al. (2005), Burgarella et al.  (2005) and Kong et al.
(2004) are plotted. We also show the well-known prescription for
starburst galaxies by Meurer et al. (1999).  The shaded region denotes the
uncertainty in our calibration that is estimated using error propagation of eq.
(\ref{eq:irxextfinal}), without considering the measurement errors in FUV and
TIR luminosities. The range of this region in IRX indicates the coverage by our
sample objects, -0.19 to 2.39 dex (see Section \ref{subsec:limitations}).
Outside this range the relation shown by the black solid line is simply the
extrapolation of that defined by our data. 

Before comparing the calibrations in the literature to our A$_{\rm FUV}$ -- IRX
relation, it is interesting to compare the published relations to each other
first. From Figure \ref{compareBuatSBresifuv.eps}, it can be clearly seen that
over the whole range of IRX probed by our sample galaxies, the relations given
by Buat et al. and Burgarella et al. are almost indistinguishable from each
other -- the difference is 0.1 mag at maximum. When compared to the Kong et al.
relation, which is also derived for normal star-forming galaxies, at low IRX
(L(TIR)$< 10{\rm L(FUV)_{obs}}$), the three relations for star-forming galaxies
stay close to each other -- the difference is no more than 0.1 mag in the
estimated attenuations.  But at higher IRX (L(TIR)$> 100{\rm L(FUV)_{obs}}$)
the Kong et al.  relation shows significant difference from the other two,
0.3-0.5 mag.  Interestingly, the difference between the starbursts relation and
the Buat et al. and Burgarella et al. relations is not dramatic (no more than
0.5 mag), given that they are derived for different types of galaxies. This
comparison tells us for galaxies with $-0.19<$ IRX $<2.39$, when the
attenuation prescriptions were used blindly, regardless of galaxy types, the
uncertainty in the UV attenuation would be less than 0.5 mag.

Since our sample galaxies are mostly normal star-forming galaxies in the local
universe (see Section \ref{subsec:limitations}), it is expected that our
calibration is more similar to those calibrated for star-forming galaxies than
for starbursts. Figure \ref{compareBuatSBresifuv.eps} confirms this
expectation. The starburst curve obtained by Meurer et al.  systematically
overestimates the A$_{\rm FUV}$ compared to our calibration, although the
difference is not dramatic, still within $1\sigma$ uncertainty of our
calibration.  The difference between starbursts and star-forming galaxies is
expected because the mean age of the dust heating population should be
systematically lower in starburst galaxies than in normal star-forming
galaxies.  Overall our calibration is very close to those for star-forming
galaxies derived by Burgarella et al. and  Buat et al.  over the whole range of
IRX spanned by our sample.  It is worth noting that the prescriptions given by
Burgarella et al. and Buat et al.  are in polynomial forms, different from the
form presented in eq.  (\ref{eq:uvextirx}), which contributes to part of the
differences. Among the three star-forming relations that we explored here, that
of Kong et al. (2004) appears to deviate the most, as seen in Figure
\ref{compareBuatSBresifuv.eps}.  At  $0 < {\rm IRX} < 1.8$ the consistency
between Kong et al. relation and ours is good, but at ${\rm IRX} < 0$ and ${\rm
IRX} > 1.8$ the relation in Kong et al. over- and under-estimate ${\rm
A_{FUV}}$ at $> 1\sigma$ level, respectively.

We compare our calibration on A$_{\rm FUV}$ versus FUV-NUV relation with those
in the literature in Figure \ref{compareAfuvuvcolorbisecresi.eps}.  Similar to
Figure \ref{compareBuatSBresifuv.eps}, the A$_{\rm FUV}$ versus FUV-NUV
relation is plotted in the top panel, the difference in A$_{\rm FUV}$ between
the prescriptions in the literature and ours is shown in the middle panel, and
the normalized (percentage) residual is shown in the bottom panel. The shaded
region denotes the uncertainty in our calibration over the range of FUV-NUV color
covered by our sample galaxies. As before the relation obtained in Section
\ref{sec:calibration} is plotted with a black solid line and the calibrations
presented by other authors are represented by different types of lines as
denoted in the top panel.  In this case, the relations plotted include the
starburst galaxy calibration by Meurer et al. (1999), the calibrations for
spectroscopically and NUV-r color selected star-forming galaxies by Salim et
al. (2007), the relation by Treyer et al. (2007) for star-forming galaxies
selected spectroscopically and the relation by Seibert et al. (2005) for a wide
assortment of galaxy types.  

It is obvious that at a given FUV-NUV color, the starburst line derived by
Meurer et al.  (1999) overestimates the A$_{\rm FUV}$ of normal star-forming
galaxies dramatically compared to others. This is consistent with the
well-known deviation of star-forming galaxies from starbursts in IRX-$\beta$
diagram (e.g., Kong et al. 2004; Cortese et al. 2006; Dale et al. 2007; Johnson
et al. 2007b). The relations defined by star-forming galaxies stay close to
each other but with different slopes and intercepts. Most of the differences
shows up at FUV-NUV $< 1$. For FUV-NUV $> 1$, the relations converge except for
those by Salim et al., which have constant A$_{\rm FUV}$ for galaxies redder
than $0.90$ and $0.95$ in FUV-NUV color for NUV-r and spectroscopically
selected galaxies respectively.  As discussed in Section \ref{sec:other}, the
calibration derived by Treyer et al.  (2007) underestimates the attenuation by
0.4 mag at FUV-NUV $\sim 0$ mag with the discrepancy decreasing as the object
gets redder in FUV-NUV color. At FUV-NUV=1.7 mag, the Treyer et al. calibration
gives consistent result with ours. The calibration by Seibert et al. (2005) is
the most consistent one with ours, and the estimated A$_{\rm FUV}$ from their
relation agrees with that from ours within $\pm1\sigma$ for the range of
FUV-NUV 0.2--1.7 mag. For bluer color (0 -- 0.2 mag), the difference is a bit
larger but always smaller than 0.2 mag in absolute scale.  In terms of relative
difference (bottom panel), both Treyer et al. and Seibert et al. relations are
consistent with ours within $\pm1\sigma$ at FUV-NUV $>$ 0.5, with increasing
discrepancy at bluer color.  For Salim et al. calibrations, both the absolute
and the relative differences are large, far beyond $\pm1\sigma$, for galaxies
redder than $\sim 0.90$ in FUV-NUV because of the plateau in A$_{\rm FUV}$
defined in these relations, which is hard to be tested using our sample because
of the limited range in FUV-NUV color covered by our galaxies (as indicated by
the shaded region).

\subsection{Comparison with Dust Attenuation Laws\label{subsec:comdustlaw}}

As shown in Section \ref{sec:calibration}, the scale parameter $s_{\rm FUV}$ in
equation (\ref{eq:tmp}) represents the slope of the UV part of the attenuation
curve. We derive $k_{\rm FUV}/k_{\rm NUV}$ from $s_{\rm FUV}$ and compare it
with those derived from the commonly used dust attenuation laws published by
Calzetti et al. (2000; originally from Calzetti et al. 1994;) and Charlot \&
Fall (2000).  The attenuation curve by Calzetti et al. (1994) was obtained
directly from the FUV-to-near-IR spectra  of a sample of UV-selected starburst
galaxies.  Charlot \& Fall (2000) used a similar but larger sample to derive
their effective attenuation curve by tuning their model parameters to account
for the distribution of the galaxies in the IRX, H$\alpha$/H$\beta$ ratio and
EW(H$\alpha$) versus $\beta$ diagrams.  The estimated values of $k_{\rm
FUV}/k_{\rm NUV}$ from our calibration, Calzetti et al. (2000) and Charlot \&
Fall (2000) are 1.35$\pm0.06$, 1.24 and 1.32, respectively, which means that in
the UV part of the dust attenuation curve, the slope of the attenuation curve
defined by our sample is consistent with that derived using Charlot \& Fall
(2000) attenuation law within the uncertainties but marginally steeper (at
$\sim 2\sigma$ level) than that estimated from Calzetti et al. (1994;2000).
The slope of our attenuation curve also suggests that the global FUV and NUV
attenuations show no evidence for the effects of a strong 2175$\AA$ bump (e.g.,
O'Donnell 1994). For example, the O'Donnell Milky Way extinction curve has a
$k_{\rm FUV}/k_{\rm NUV}$ ratio close to 0.9, which is much smaller than
1.35$\pm0.06$ defined by our galaxies.

Based on the attenuation in the FUV estimated using our prescriptions, we can
study the A$_{\rm FUV}$ -- A$_{{\rm H}\alpha}$ relation. Figure
\ref{Afuv.Aha.Cal.eps} shows the IRX-derived (left panel) and FUV-NUV derived
(right panel) A$_{\rm FUV}$ as a function of H$\alpha$/H$\beta$-derived
A$_{{\rm H}\alpha}$. The solid line in this figure is derived from eqs.
(\ref{eq:Ahafnc}) and (\ref{eq:fncextfinal}), which present the relations
between FUV-NUV color and A$_{\rm H\alpha}$ and A$_{\rm FUV}$ respectively.
For comparison,  the relation derived from Calzetti (2001) law after adapted to
GALEX FUV waveband, which has the form ${\rm A_{FUV}}=1.82{\rm A_{H\alpha}}$,
is plotted as the dotted line.  As can be seen from the figure, at a given
H$\alpha$ attenuation, Calzetti's law gives a lower FUV attenuation estimate
than ours and the difference becomes more significant for galaxies that are
more attenuated.  Quantitatively speaking, however, the slope of the solid
line, which determined from our data, is 2.06$\pm$0.28, consistent with that
given by Calzetti's law within the error.

\subsection{Comparisons Made For NUV}

We performed similar comparisons for NUV with the calibrations published in
the literature.  When compared to the calibrations in the literature, similar to
the results we saw for FUV, our calibration is in good agreement with the one
given by Burgarella et al. (2005) within 0.1 mag for the whole range of
Log[(TIR)/L(NUV)]. The A$_{\rm NUV}$/A$_{\rm H\alpha}$ ratio defined by our
data is 1.52$\pm$0.56, coincident with the predicted value from Calzetti
(2001)'s law within the error.

\subsection{Dust Heating From Old Stars\label{subsec:oldpops}}

It is now well-known that a significant fraction of IR emission comes from dust
heating by old stellar populations (e.g., Sauvage \& Thuan 1992; Bell 2003 and
references therein). In Paper I, by comparing our calibration for H$\alpha$+TIR
as a SFR indicator with stellar population synthesis models, we found that
for our sample objects up to 50\% of the TIR emission could be from stars older
than 100 Myr. Are the results presented here consistent with those estimates?

As mentioned in Section \ref{sec:calibration}, the coefficient $a_{\rm FUV}$ in
eq.  (\ref{eq:uvextirx}) is the product of the inverse of the bolometric
correction $\eta_{\rm FUV}$ and the factor ${(1 - {\rm e}^{-\tau_{\rm FUV}})}
\over {(1 - {\rm e}^{-\overline{\tau}})}$.  In Paper I, we have empirically
derived $\bar{\tau}$ by combining the TIR luminosity and the multi-wavelength
SEDs of a subset of MK06 galaxies with GALEX, SDSS and 2MASS observations.
Briefly, eq. (\ref{eq:bolometric}) was used to estimate $\bar{\tau}$ from the
TIR and the bolometric luminosities, which were derived by adding the TIR
luminosity to the integrated luminosity from the FUV to K band.  After
substituting the values of $\tau_{\rm FUV}$ and $\bar{\tau}$ into  ${(1 - {\rm
e}^{-\tau_{\rm FUV} })} \over {(1 - {\rm e}^{-\bar{\tau}})}$, we find that the
median of ${(1 - {\rm e}^{-\tau_{\rm FUV} })} \over {(1 - {\rm
e}^{-\bar{\tau}})}$ is around 1.5. Then the bolometric correction $1/\eta_{\rm
FUV}$ is estimated to be $\sim 3.3$ by comparing $a_{\rm FUV}$ with ${(1 - {\rm
e}^{-\tau_{\rm FUV} })} \over {(1 - {\rm e}^{-\bar{\tau}})}$.

In order to estimate the effect of dust heating from old stars, we compare the
bolometric correction derived from our data to those from stellar population
synthesis models.  We found that the bolometric correction defined by our data
is more than two times greater than the model prediction for a 100 Myr old
stellar population with constant star formation history.  This implies that
stars older than 100 Myr must contribute to the dust heating.  The spectral
synthesis modelling using STARBURST99 indicates that the bolometric correction
for a 10 Gyr old stellar population with constant star formation history, solar
metallicity and either a Salpeter or a Kroupa IMF is only 15\%--20\% lower than
the estimated value above. In other words, the mean property of our sample
objects approximates a stellar population that has been constantly forming
stars for 10 Gyr, which is exactly what we inferred from the analyses of
H$\alpha$+TIR in Paper I. It is worth noting that the consistent inference
drawn from Paper I and this paper ensure us the robustness of our results.
However, this estimate of the average stellar population of our sample galaxies
conflicts with the dust-free FUV-NUV color derived in Section
\ref{sec:calibration}, which is $\sim 0.04$ mag redder than a 10 Gyr old galaxy
with constant star formation history.  All the estimates made here are very
rough and they should only be used as a guide to our understanding of the
underlying physics shown by the data qualitatively.  More careful modelling is
needed for a quantitative understanding.

\subsection{Systematic Uncertainties\label{subsec:sys}}

As discussed in Section \ref{subsec:limitations}, our calibrations were derived
over limited ranges of galaxy properties. So it is important to understand possible
systematics which were built into our prescriptions. 

First of all, prescriptions calibrated for different types of galaxies are
different.  This is not surprising because different types of galaxies show
different IRX-$\beta$ (i.e., FUV-NUV color) relations. As shown in Section
\ref{subsec:comparisons}, the IRX-based and the FUV-NUV color-based attenuation
correction prescriptions for star-forming galaxies differ from those for
starburst galaxies. Another two extreme examples are dwarf irregulars and
(Ultra)Luminous Infrared Galaxies ((U)LIRGs).  Dale et al. (2009) examined the
location of the Local Volume Legacy (LVL) sample galaxies in the IRX-$\beta$
diagram. Nearly two-thirds of LVL galaxies are dwarf/irregular systems.  Those
authors showed that the majority of LVL galaxies fall below the IRX-$\beta$
relation as defined by normal star-forming galaxies, and they cluster in an
area with blue FUV-NUV color and low IRX (see also Lee et al. 2009). The range
in IRX spanned by the LVL galaxies, for a given FUV-NUV color, is large -- more
than an order of magnitude.  For (U)LIRGs, Howell et al. (2010; see also
Goldader et al. 2002; Calzetti 2001) showed that (U)LIRGs in the Great
Observatories ALL-sky LIRG Survey (GOALS) lie above the starburst relation, and
the correlation between IRX and FUV-NUV color is not strong either.  Similar to
dwarf/irregulars, at a given FUV-NUV color, the (U)LIRGs span more than an
order of magnitude in IRX.  One assumption we made in our calibration method is
that the attenuation derived from IRX matches that estimated from FUV-NUV
color, which is reasonable for our normal star-forming galaxy sample given the
relatively tight correlation between IRX and FUV-NUV color, as shown in Figure
\ref{irxfuv.uvcolor.eps}. However, it breaks down for dwarf irregulars and
(Ultra)Luminous Infrared Galaxies given the large range of IRX spanned by these
galaxies at a given FUV-NUV color. Therefore no attempt was made to calibrate
an IRX-based and FUV-NUV color-based attenuation correction prescriptions for
these galaxies. 

Apart from different galaxy samples, the differences between different
calibrations can be traced to the different methodologies.  Previous
calibrations have been based either on pure model assumptions, or on a
comparison of observational data with models. The difference in the form of the
parameterization of the prescriptions also contributes to the differences.  As
we can see from the comparisons in Section \ref{subsec:comparisons}, the
differences in methodology and parameterization (Buat et al. 2005; Burgarella
et al. 2005 and this work) only account for a small portion of the differences,
which cannot even be distinguished within $\pm1\sigma$ uncertainty, and the
main contributors to the different relations are the star formation history and
the dust attenuation curve that includes the effect of the dust/star
distribution, as suggested by other authors (e.g., Charlot \& Fall 2000; Kong
et al. 2004; Burgarella et al.  2005).  

It is difficult to test the impact of dust attenuation curves on attenuation
calibrations. However, it is possible to estimate the effect of dust-heating
stellar populations.  As discussed in Section \ref{subsec:oldpops}, for our
sample galaxies, up to half of the TIR emission could be from dust heated by
stars older than 100 Myr.  By contrast, in starburst galaxies, the dust is
dominantly heated by stellar populations younger than 100 Myr (Meurer et al.
1999). This difference in dust-heating stellar populations of starbursts and
our star-forming galaxies results in the difference in attenuation estimates,
0.3 mag at maximum over $-0.19<$ IRX $<2.39$, as shown in Figure
\ref{compareBuatSBresifuv.eps}.  The effects on the UV color-based corrections
are much larger than on IRX-based prescriptions.  The IRX-based prescriptions
are surprisingly consistent with each other.  However different FUV-NUV based
attenuation schemes have by comparison enormous inconsistencies.  This
underscores our concerns about color-based attenuation corrections based on
scatter in our own data.

In principle, by comparing modelled SEDs with multi-wavelength data, the
differences in star formation histories and dust attenuation curves can be
taken into account on a galaxy-to-galaxy basis. However, the dust attenuations
derived this way and the resulting relations of ${\rm A_{FUV}}$ versus IRX or
${\rm A_{FUV}}$ versus FUV-NUV color are dependent on the adopted stellar
population synthesis models and the dust attenuation curves. On the other hand,
the requirement of the availability of large libraries of models and
multi-wavelength data makes this method less practical in many cases.  

By contrast, our approach is more straightforward and less model-dependent. Our
calibrations rely almost entirely on the data themselves and hence can serve as
independent checks of other model-dependent methods. Although we assumed an
average stellar population and dust attenuation curve (as discussed in Section
\ref{sec:calibration}), they do not introduce systematics into the calibrations
and are valid on a statistical basis.  Therefore, our calibrations should be
robust. They also tend to be consistent with most of the published calibrations
for normal star-forming galaxies.

The last possible systematic error comes from the TIR luminosities we adopted
in our calibrations. We used TIR luminosities based on IRAS observations, which
are systematically lower than those based on Spitzer/MIPS measurements by $24\%
\pm 23\%$ on average (Section \ref{sec:sample} and Figure 1 in Paper I). If we
simply substitute the IRAS TIR luminosities by the MIPS TIR luminosities, the
coefficient in front of IRX in eq. (\ref{eq:irxextfinal}) will be lower by
24\%, i.e., changes to 0.37, which is within the $1\sigma$ uncertainty of the
coefficient.  So if one uses MIPS TIR luminosities to estimate the FUV
attenuation using our IRX-based prescription (i.e., eq. (\ref{eq:irxextfinal})),
the uncertainty will be less than 0.3 mag.

In summary, our IRX-based attenuation correction has an uncertainty of $< 0.3$
mag, when it is applied to starburst galaxies or to galaxies with MIPS based
TIR luminosities.  By contrast, if one uses our FUV-NUV color prescription to
estimate the FUV attenuation for starbursts, he can underestimate the FUV
attenuation by more than 1-2 mag. But for dwarf irregulars or (U)LIRGs, our
prescriptions may suffer from severe problems.

\subsection{The SFR Matching Method}

In the introduction, we mentioned that Treyer et al. (2007) derived an A$_{\rm
FUV}$ -- FUV-NUV relation by matching the FUV-based and H$\alpha$-based SFRs.
The attenuation derived this way is dependent on the intrinsic ratio of the FUV
to H$\alpha$ luminosity, which in turn is dependent on model assumptions. Here
we use a similar method to re-calibrate the A$_{\rm FUV}$--IRX relation in
order to evaluate the model dependence of the calibration obtained this way.
We converted the Balmer-corrected H$\alpha$ luminosity to a Balmer-corrected
FUV luminosity using a theoretical FUV/H$\alpha$ ratio and then force the
combined observed (i.e., without internal dust attenuation correction) FUV and
TIR luminosity to match the Balmer-corrected FUV luminosity so that the
coefficient $a_{\rm FUV}$ in eq. (\ref{eq:uvextirx}) can be obtained.  We used
the same set of models as those used in Figure \ref{comlumiirx.his.eps} and
Figure \ref{comlumifnc.his.eps}.  The model predictions and the best-fit
coefficients $a_{\rm FUV}$ are listed in Table 2 (For reference the
corresponding quantities for NUV are also listed.). Consistent with what we saw
from Figure \ref{comlumiirx.his.eps}, Table 2 shows that our calibration
derived in Section \ref{sec:calibration} is in excellent agreement with that
predicted by assuming a constant star formation history with Kroupa IMF and age
100 Myr. 

The FUV/H$\alpha$ ratio changes at most by 10\% when only the effect of a
single parameter is taken into account, and accordingly the resulting
coefficient $a_{\rm FUV}$ varies by 11\%.  The change in the coefficient
directly affects the estimated attenuation from IRX.  This impact is shown in
Figure \ref{compareAfuvirxdiffIMFage.eps}.  It is clear that assuming a
Salpeter or a Kroupa IMF does not affect the calibrations dramatically -- at
maximum $\sim$ 0.07 mag over the range of IRX=0 -- 3.  The calibrations are
also affected by assuming either a 100 Myr old or a 1 Gyr (or older) old
stellar population -- at maximum $\sim$ 0.11 mag over the range tested. However
the most widely used K98's prescriptions give larger discrepancy -- -0.2 mag at
maximum when IRX=3. This large discrepancy with K98's prescriptions is probably
due to the use of the old stellar evolutionary models, which leads to an
over-estimate of H$\alpha$ luminosity by 11\% compared to that from STARBURST99
under the same set of assumptions of IMF, metallicity and star formation
history.

From these tests, we can draw the following conclusions:
(1) The calibration established using our new method is consistent with that
based on SFR matching method for a constant star formation history with Kroupa
IMF and age 100 Myr. (2) Compared to our prescription, the adoption of a Salpeter IMF
and/or an older age both leads to an overestimate in A$_{\rm FUV}$ by up to 0.1
mag over the range of IRX=0 -- 3.  (3) Relative to our calibration, the widely
used K98's SFR prescriptions can underestimate A$_{\rm FUV}$ by up to 0.2 mag
(at IRX=3).

\section{SUMMARY \label{sec:summary}}

We have calibrated $\log {\rm [L(TIR)/L(FUV)_{obs}]}$ (i.e., IRX) and the
FUV-NUV color as FUV dust attenuation indicators using a nearby star-forming
galaxy sample studied by Moustakas \& Kennicutt (2006).  The combinations of
25\um\ and 1.4GHz radio continuum luminosities with the observed FUV
luminosities were also empirically calibrated to probe the dust attenuation
corrected FUV luminosities. Similar calibrations were derived for NUV band as
well.  The coefficients in these calibrations are summarized in Table 3 (see
Section \ref{sec:uncertainties}). These prescriptions provided in Table 3 can
be used to derive attenuation-corrected star formation rates. Our main results
can be summarized as follows.

1. The IRX-corrected FUV luminosities based on our new calibration, i.e., 
$${\rm L(FUV)_{corr}=L(FUV)_{obs}}+(0.46\pm0.12) {\rm L(\mbox{TIR})},$$
show tight and linear correlation with the Balmer decrement corrected H$\alpha$
luminosities, with a rms scatter of $\pm$0.09 dex ($\pm 0.23$ mag).
Statistically speaking, their ratios are consistent with the model predictions
from STARBURST99 by assuming a constant star formation rate over 100 Myr and
solar metallicity.  This consistency applies whether a Salpeter or a Kroupa IMF
with lower and upper mass limits of 0.1 and 100 \msun\ is adopted.
Furthermore, these dust attenuation corrected FUV/H$\alpha$ ratios do not show
any trend with other galactic properties over the ranges covered by our sample
objects.  These results suggest that linear combinations of TIR luminosities
and the observed FUV luminosities (without internal attenuation corrections)
are excellent star formation rate tracers.

2. The FUV-NUV corrected FUV luminosities are broadly correlated with the
Balmer-corrected H$\alpha$ luminosities. But they do not trace each other
linearly and the rms scatter is large -- $\pm$0.23 dex ($\pm 0.58$ mag), which
is $\sim 2.5$ times larger than the case for IRX-corrected FUV luminosities. In
addition, the attenuation corrected FUV/H$\alpha$ ratios show correlations with
a few other galactic properties, which will introduce systematic errors into
the attenuation estimates in different galaxy samples.  We confirm others'
findings that FUV-NUV color is not a good dust attenuation indicator for normal
star forming galaxies though it can be used with caution when IR data are not
available.

3. Linear combinations of 25\um\ and 1.4GHz radio continuum luminosities
with the observed FUV luminosities can be used as surrogates of IRX-corrected
FUV luminosities to trace the attenuation-corrected star formation rates. Their
correlations with the Balmer-corrected H$\alpha$ luminosities are slightly less
tight than the IRX-corrected FUV luminosities -- 0.13 dex and 0.14 dex, and
a non-linearity is shown in the correlation between FUV+1.4GHz and
Balmer-corrected H$\alpha$ luminosities with a slope of $1.09\pm0.02$.

4. Our calibrations are tested on normal star-forming galaxies with
Balmer-based A$_{{\rm H}\alpha}$ = 0 -- 2.5 mag, IRX = $-0.19$ -- 2.39 and
FUV-NUV = 0.07 -- 1.06 mag.  The A$_{\rm FUV}$ -- IRX relation shows overall
consistency with prescriptions for star-forming galaxies by Burgarella et al.
(2005) and Buat et al. (2005) within $\pm1\sigma$ uncertainty.  While our
A$_{\rm FUV}$ -- FUV-NUV color relation is in good agreement with that
presented by Seibert et al. (2005). When our IRX-based attenuation correction
is applied to starburst galaxies or to galaxies with MIPS based TIR
luminosities, it has an uncertainty of $< 0.3$ mag. By contrast, if our FUV-NUV
color prescription is used to estimate the FUV attenuation for starbursts, it
can underestimate the FUV attenuation by more than 1-2 mag.

5. The estimated A$_{\rm FUV}$ using our prescriptions is related to the Balmer-based 
A$_{\rm H\alpha}$ in a way consistent with that given by Calzetti (2001)'s law within
the error.

6. Our analyses show that the combination of the NUV and the TIR luminosities
can serve as a rather good SFR indicator for our sample galaxies, but may
become problematic in galaxies with non-constant star formation histories in
the past $\sim 1$ Gyr.

\acknowledgements

C.-N. Hao acknowledges the support of a Royal Society UK--China Fellowship.
C.-N. Hao also acknowledges the support from the NSFC (Grant No.  10833006 and
11003015).  We thank an anonymous referee for helpful comments that improved the
paper.  Some/all of the data presented in this paper were obtained from the
Multimission Archive at the Space Telescope Science Institute (MAST). STScI is
operated by the Association of Universities for Research in Astronomy, Inc.,
under NASA contract NAS5-26555. Support for MAST for non-HST data is provided
by the NASA Office of Space Science via grant NAG5-7584 and by other grants and
contracts.

\pagebreak

\begin{figure} 
\epsscale{} 
\plotone{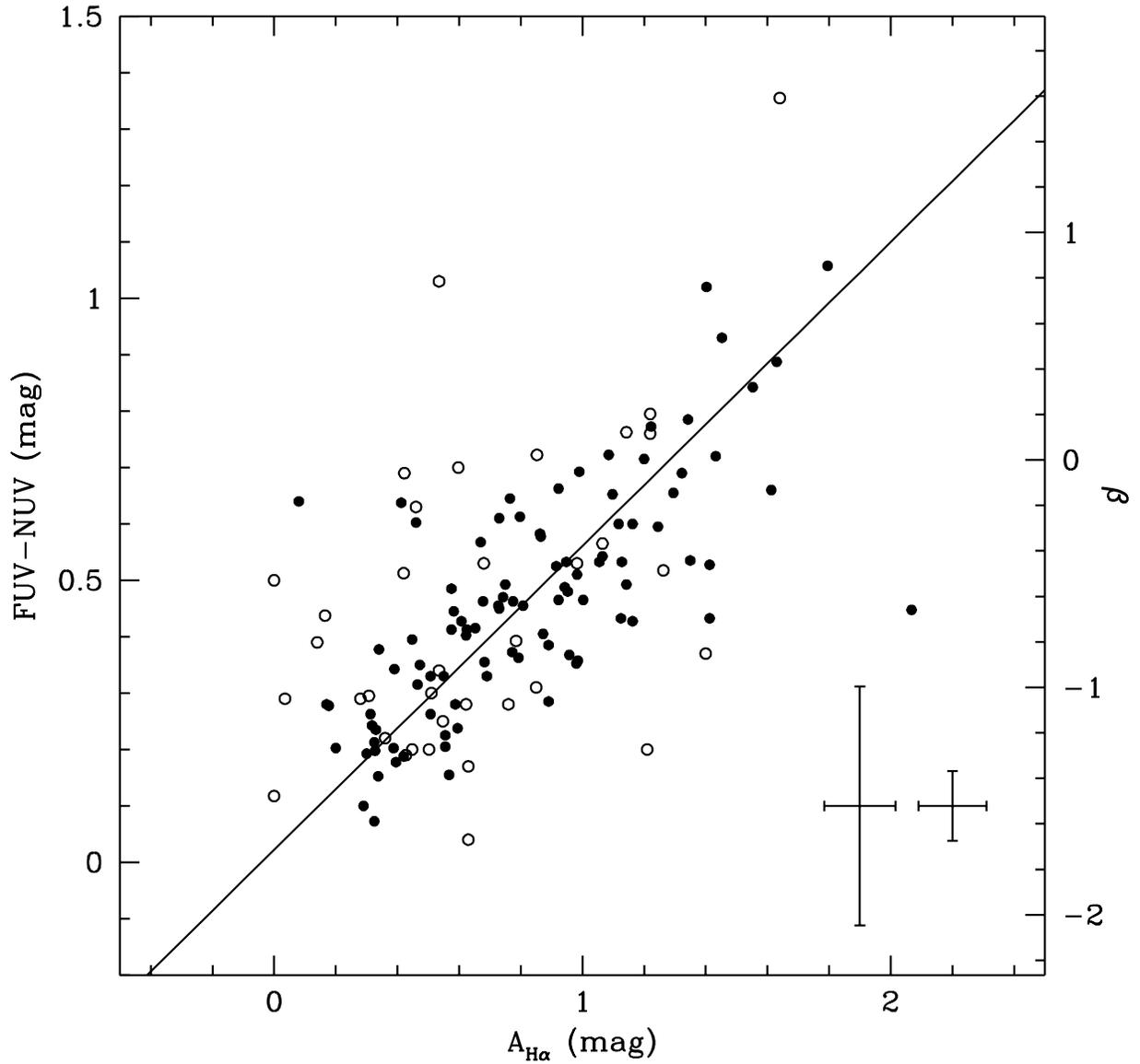}
\caption{FUV-NUV color as a function of attenuation in H$\alpha$ line derived
from the Balmer decrement ratio for MK06 sample (solid circles) and SINGS
sample (open circles), with the corresponding UV continuum slope $\beta$
labelled on the right axis.  The solid line is the bisector fitting to the MK06
sample. The error bars in the bottom-right corner denote median errors for the MK06
sample (right) and the SINGS sample (left).} 
\label{uvcolor.Aha.eps} 
\end{figure}

\begin{figure} 
\epsscale{} 
\plotone{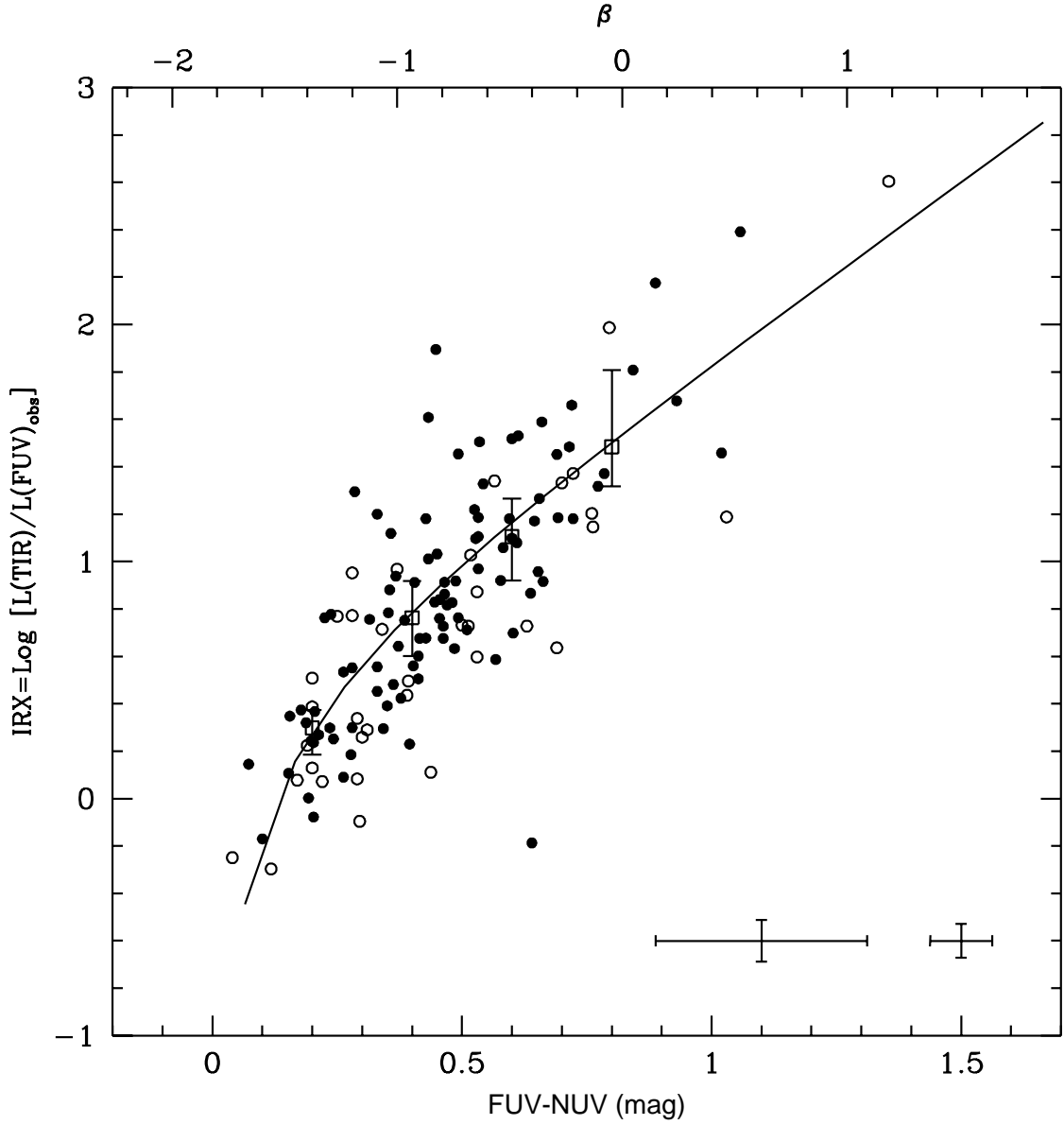}
\caption{IRX (i.e. $\log {\rm [L(TIR)/L(FUV)]}$) versus FUV-NUV color for the MK06 sample 
(solid circles) and the SINGS sample (open circles), with $\beta$ labelled on the top axis. 
The open squares with error bars represent the median, lower (25\%) and upper (75\%) 
quartiles in bins of width 0.2 mag in FUV-NUV color. The solid line, which has the form
defined in eq. (\ref{eq:irxbeta}), is the best fit to the MK06 sample. The error bars in 
the bottom-right corner denote median errors for the MK06
sample (right) and the SINGS sample (left).} 
\label{irxfuv.uvcolor.eps} 
\end{figure}

\begin{figure}
\epsscale{}
\plotone{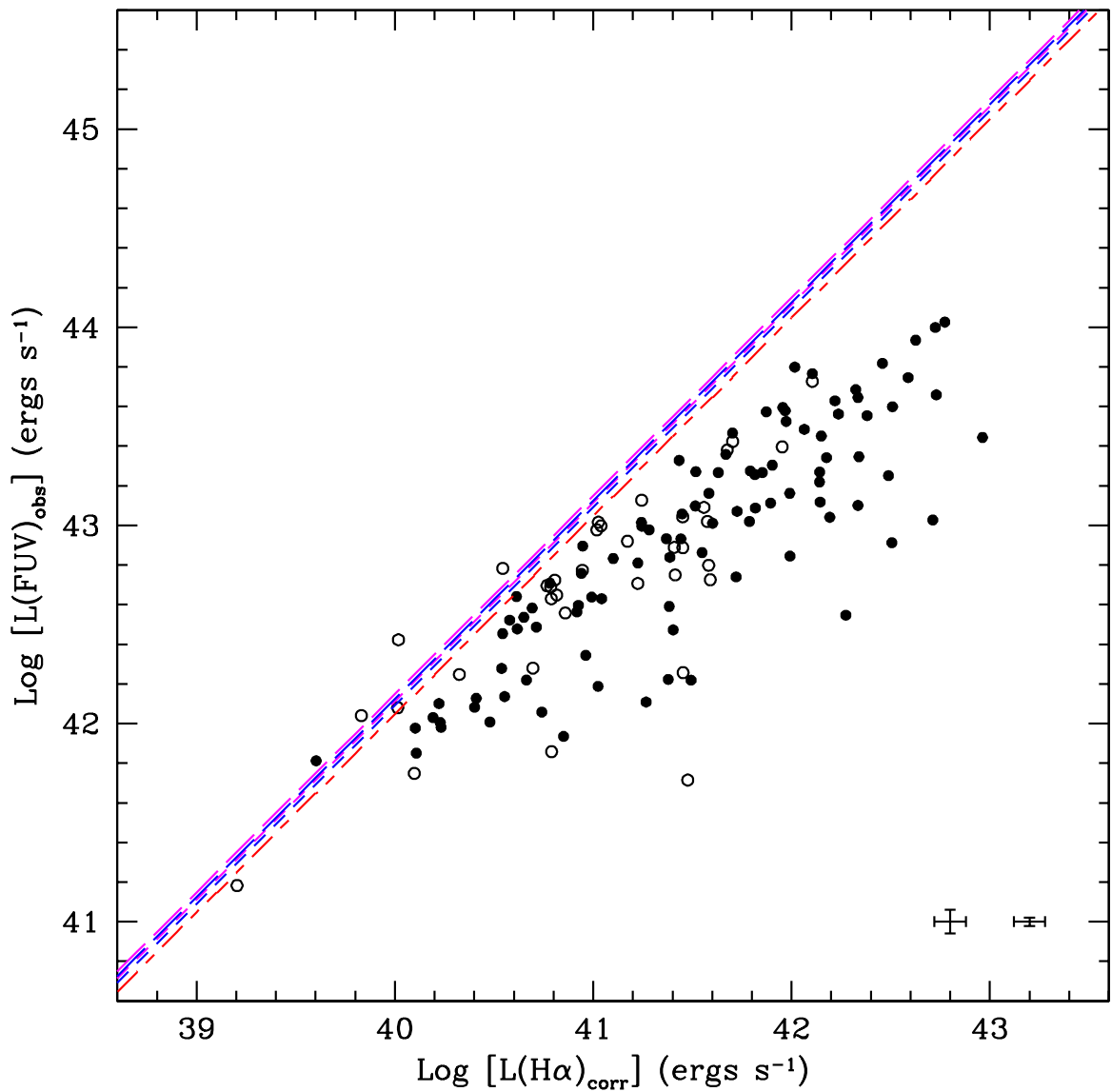}
\caption{Observed FUV luminosities without correction for internal dust
attenuation versus Balmer decrement ratio corrected H$\alpha$ luminosities for
MK06 sample (solid circles) and SINGS sample (open circles). The color-coded
lines represent the predicted relations (in the absence of dust attenuation in
the UV) by different SFR prescriptions, same as Figure
\ref{comlumiirx.his.eps}.  The error bars in the bottom-right corner denote
median errors for the MK06 sample (right) and the SINGS sample (left).}
\label{lfuvobs.lhacor.eps}\end{figure}

\begin{figure} 
\epsscale{} 
\plotone{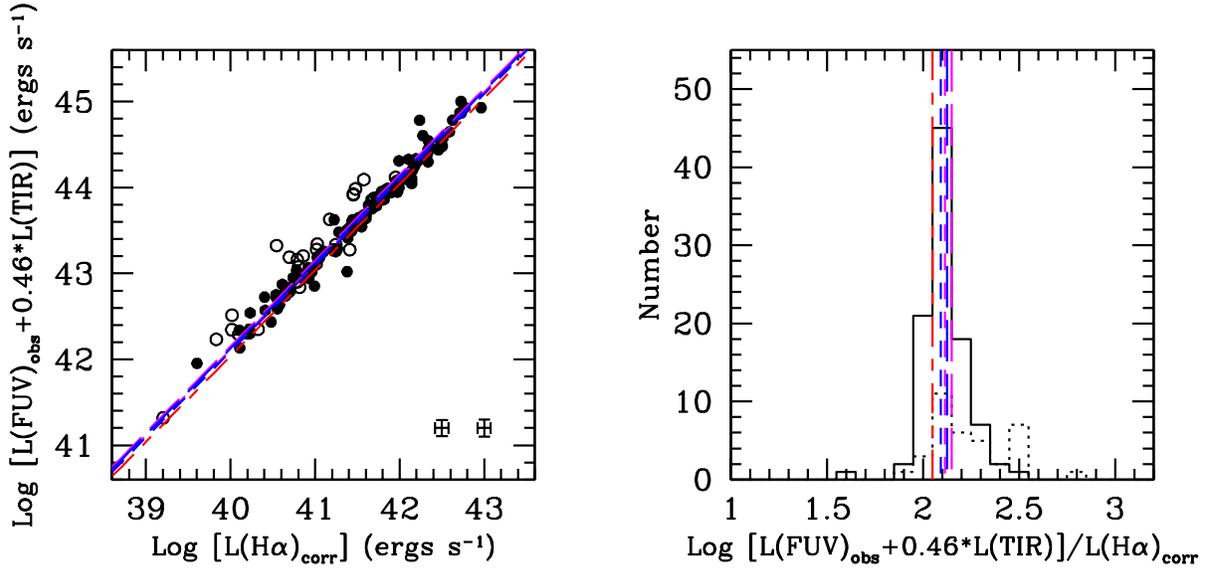}
\caption{{\it left panel:} IRX-corrected FUV luminosities as a function of
Balmer decrement ratio corrected H$\alpha$ luminosities for MK06 sample (solid
circles) and SINGS sample (open circles).  The lines overplotted represent the
predicted relations by different SFR prescriptions (see text), which are the
K98 prescription (red short-long dashed line) and models constructed using
STARBURST99 by assuming a constant star formation history, solar metallicity
for Kroupa IMF at 100 Myr (blue short dashed line), Kroupa IMF at 1 Gyr (blue
long dashed line), Salpeter IMF at 100 Myr (magenta short dashed line) and
Salpeter IMF at 1 Gyr (magenta long dashed line). The error bars in the
bottom-right corner denote median errors for the MK06 sample (right) and the
SINGS sample (left).  {\it right panel:} Histograms of the ratios of the
IRX-corrected FUV luminosities to Balmer-corrected H$\alpha$ luminosities for
MK06 sample (solid line) and SINGS sample (dotted line) in log scale. The
vertical lines represent the predicted values by different SFR prescriptions,
as shown in left panel. This panel is given to show the differences between the
models more clearly.} \label{comlumiirx.his.eps} \end{figure}

\begin{figure} 
\epsscale{} 
\plotone{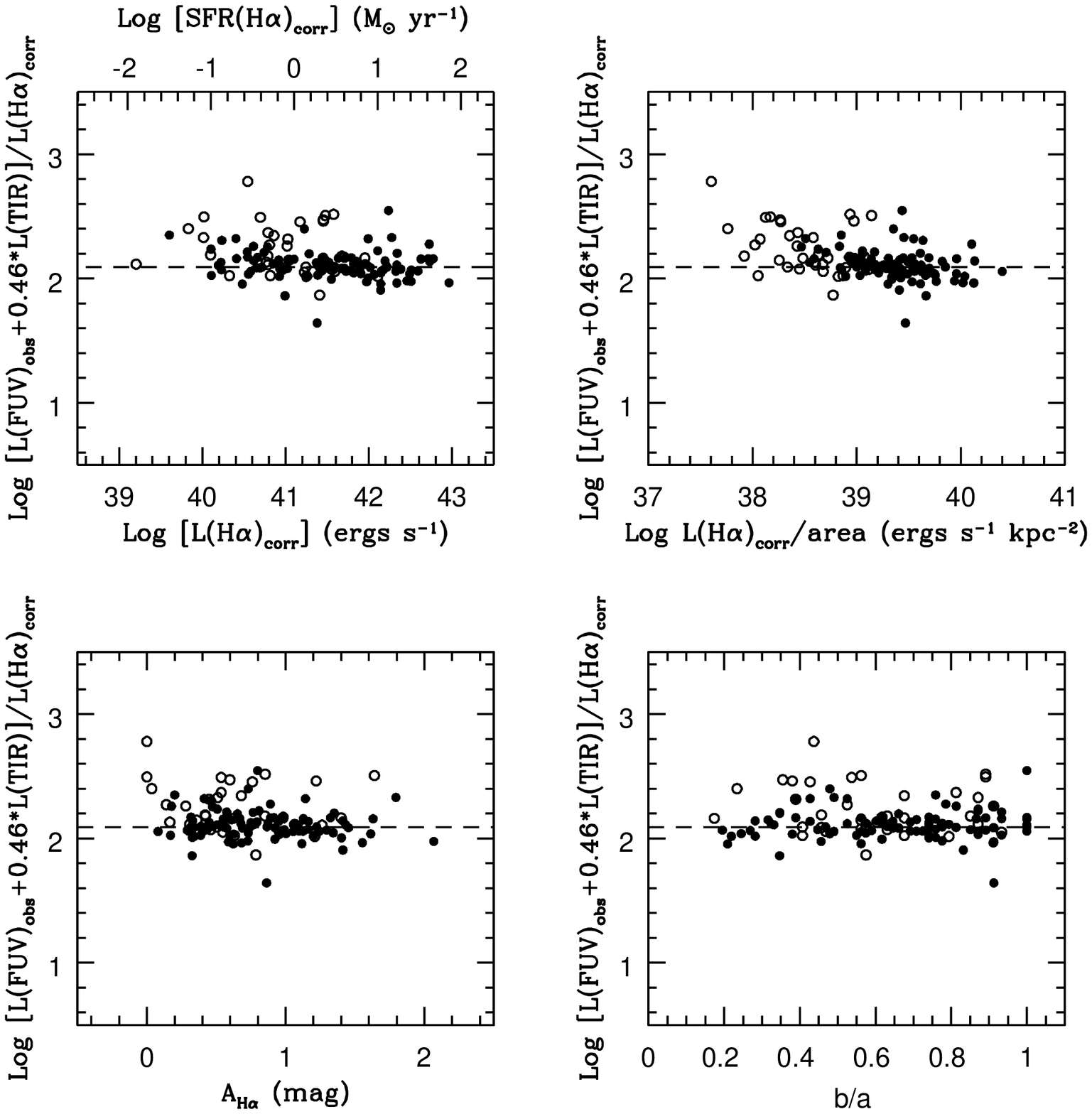} 
\caption{The logarithm residuals of the attenuation-corrected FUV luminosities
from IRX relative to the Balmer-attenuation-corrected H$\alpha$ luminosities as
functions of Balmer-attenuation-corrected H$\alpha$ luminosities (top-left
panel), Balmer-attenuation-corrected H$\alpha$ luminosities per unit area
(top-right panel), attenuation in H$\alpha$ calculated from H$\alpha$/H$\beta$
ratio (bottom-left) and axial ratio (inclination) b/a (bottom-right).  The
label at the top of the top-left panel shows the corresponding SFR according to
the L(H$\alpha$) -- SFR relation given by K98. The solid circles represent MK06
galaxies while the open circles denote SINGS galaxies. The dashed line denotes
the predicted value by the STARBURST99 synthesis model for a constant star
formation history, solar metallicity and Kroupa IMF at 100 Myr.}
\label{fuvirxresi.para1.eps} \end{figure}

\begin{figure} 
\epsscale{} 
\plotone{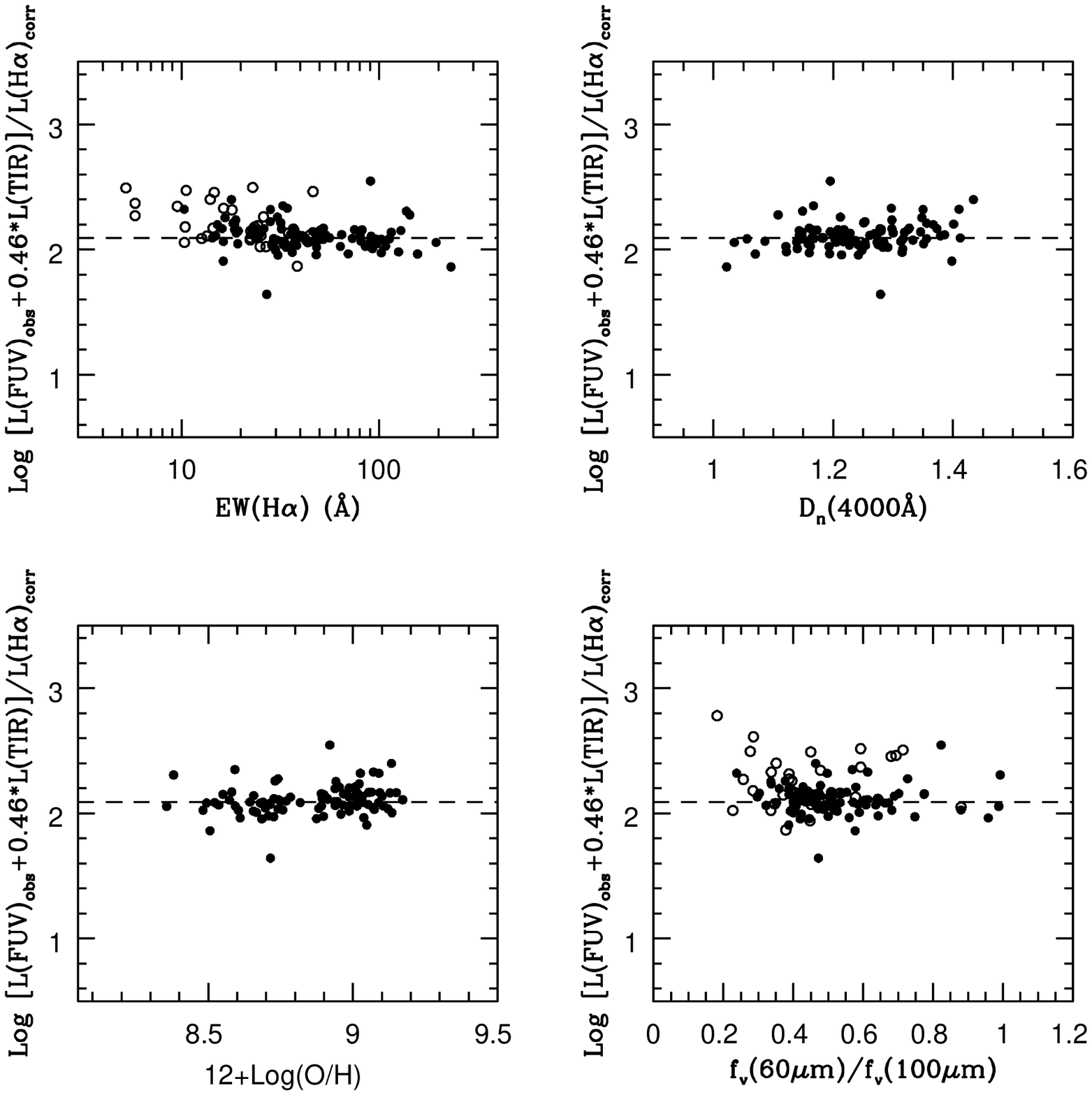}
\caption{Similar to Figure \ref{fuvirxresi.para1.eps} but with residuals
plotted as functions of integrated Equivalent Width at H$\alpha$ (top-left
panel), 4000$\AA$ break (D$_{\rm n}(4000\AA)$) (top-right panel), gas-phase
oxygen abundance (12+log(O/H)) (bottom-left panel) and FIR color
($f_\nu(60\micron)/f_\nu(100\micron)$) (bottom-right panel). See Figure
\ref{fuvirxresi.para1.eps} for the explanation of the symbols.}
\label{fuvirxresi.para2.eps} \end{figure}

\clearpage

\begin{figure}
\epsscale{}
\plotone{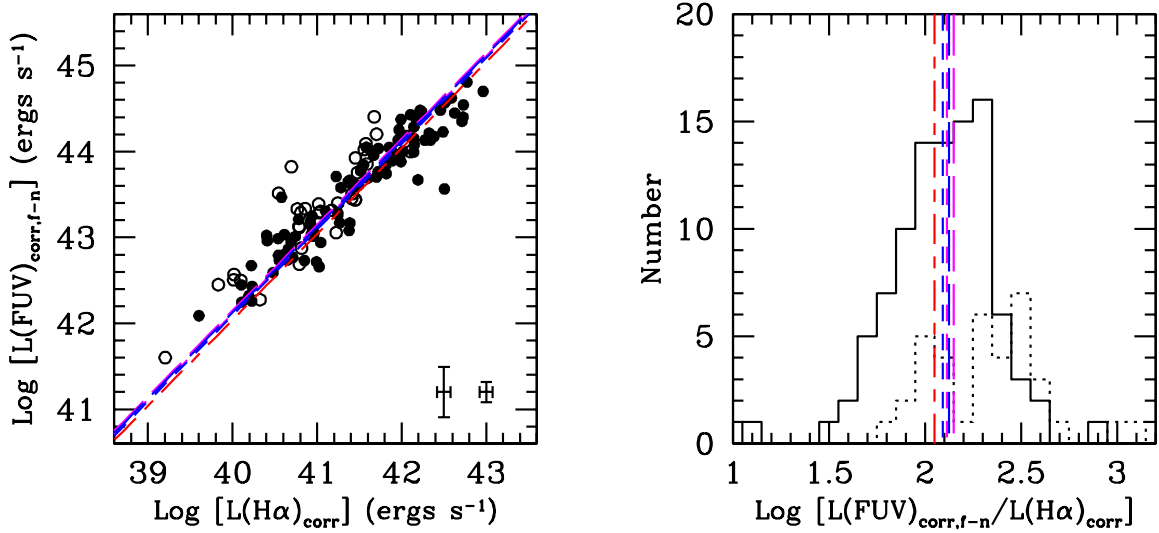}
\caption{{\it left panel:} FUV-NUV color-corrected FUV luminosities as a function of
Balmer decrement ratio corrected H$\alpha$ luminosities for MK06 sample (solid
circles) and SINGS sample (open circles).  The lines overplotted represent the
predicted relations by different SFR prescriptions, same as those in Figure
\ref{comlumiirx.his.eps}. The error bars in the 
bottom-right corner denote median errors for the MK06 sample (right) and the
SINGS sample (left). {\it right panel:} Histograms of the ratios of the FUV-NUV
color-corrected FUV luminosities to Balmer-corrected H$\alpha$ luminosities for
MK06 sample (solid line) and SINGS sample (dotted line) in log scale. The
vertical lines represent the predicted values by different SFR prescriptions,
as shown in left panel. This panel is to show the differences between the models
more clearly.} \label{comlumifnc.his.eps} \end{figure}

\begin{figure}
\epsscale{}
\plotone{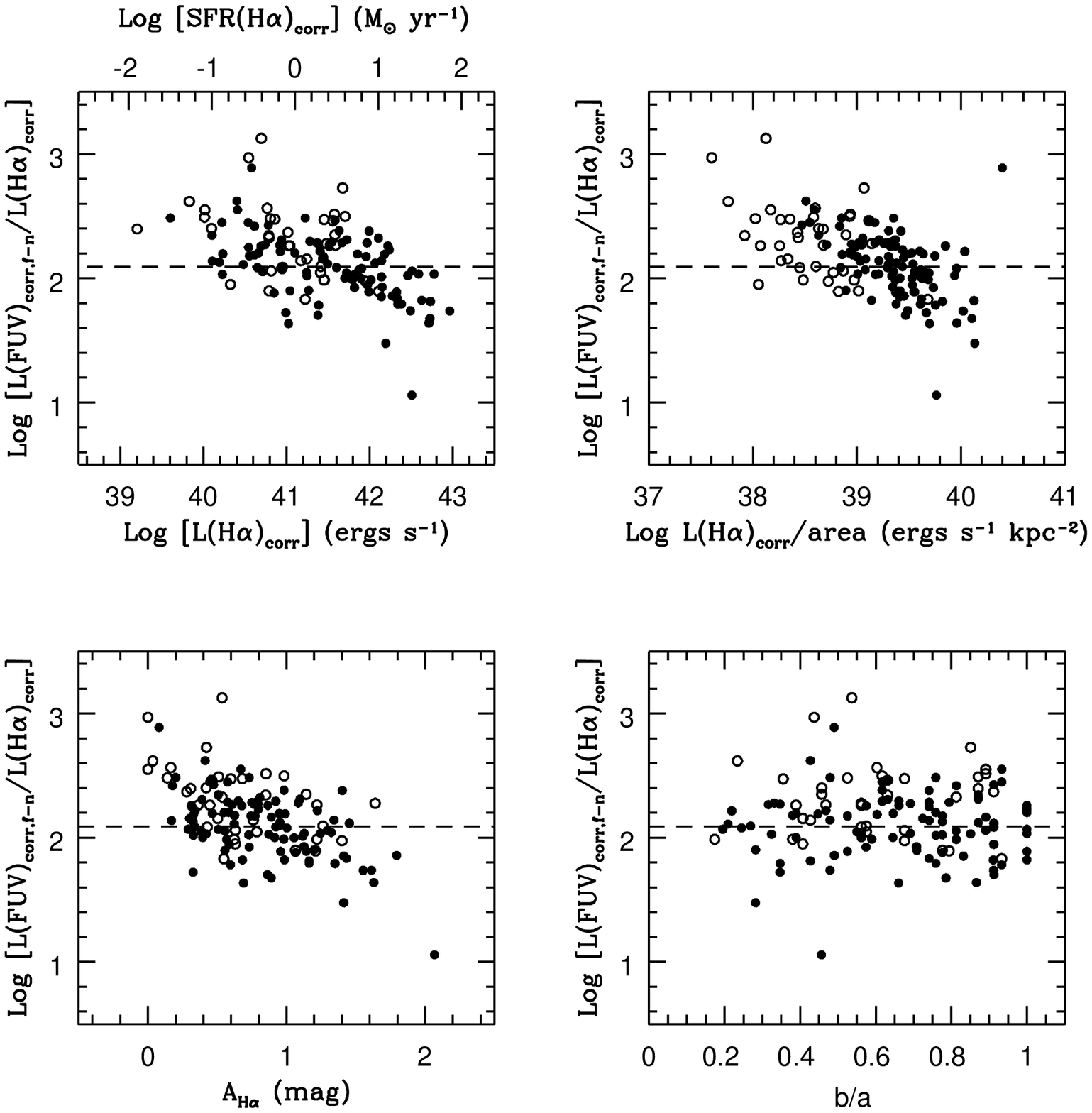}
\caption{The logarithm residuals of the attenuation-corrected FUV luminosities
from FUV-NUV color relative to the Balmer-attenuation-corrected H$\alpha$ luminosities as
functions of Balmer-attenuation-corrected H$\alpha$ luminosities (top-left
panel), Balmer-attenuation-corrected H$\alpha$ luminosities per unit area
(top-right panel), attenuation in H$\alpha$ calculated from H$\alpha$/H$\beta$
ratio (bottom-left) and axial ratio (inclination) b/a (bottom-right).  The
label at the top of the top-left panel shows the corresponding SFR according to
the L(H$\alpha$) -- SFR relation given by K98. The solid circles represent MK06
galaxies while the open circles denote SINGS galaxies. The dashed line denotes
the predicted value by the STARBURST99 synthesis model for a constant star
formation history, solar metallicity and Kroupa IMF at 100 Myr.}
\label{fuvuvcolorresi.para1.eps} \end{figure}

\begin{figure} 
\epsscale{} 
\plotone{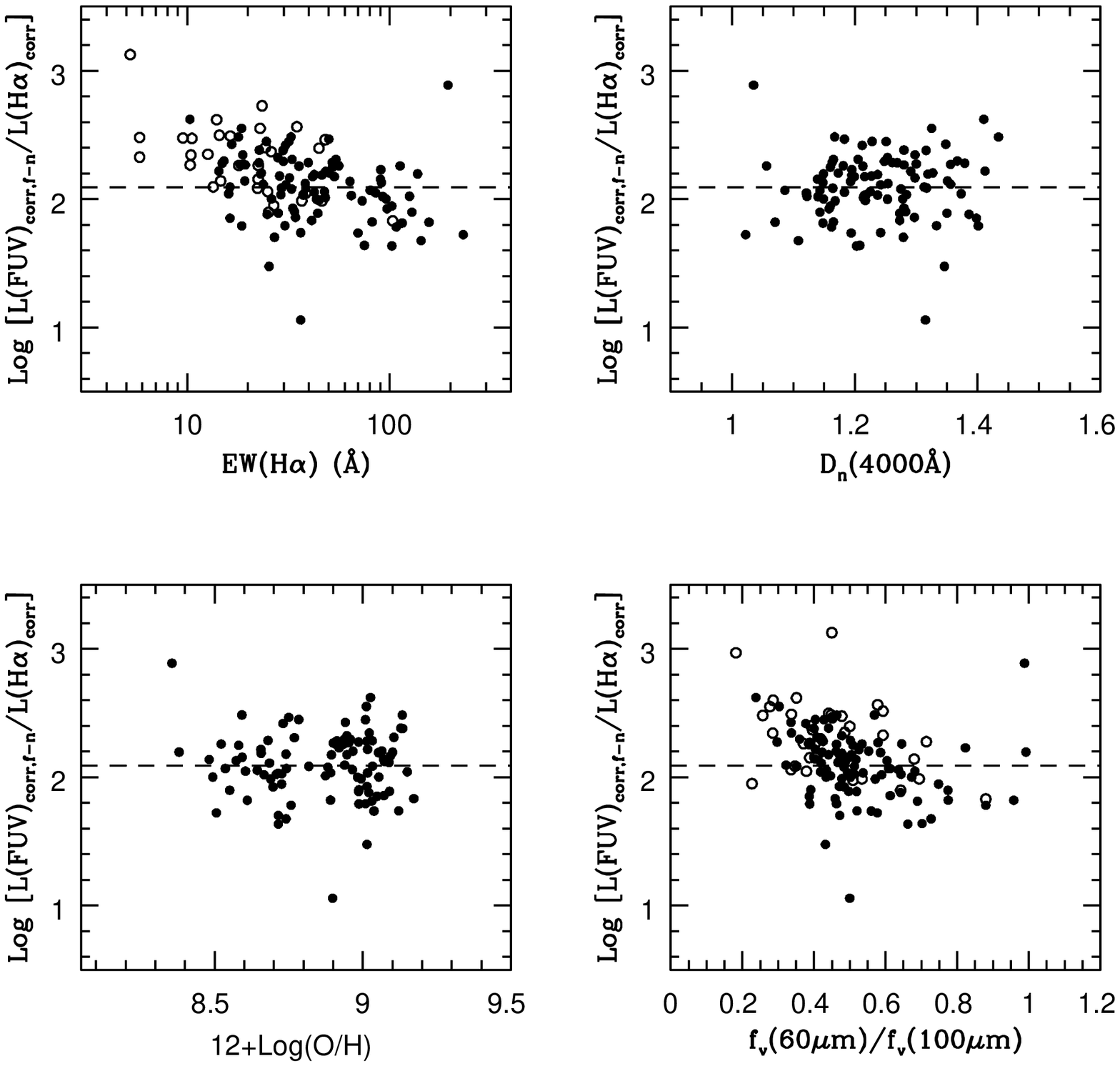}
\caption{Similar to Figure \ref{fuvuvcolorresi.para1.eps} but with residuals
plotted as functions of integrated Equivalent Width at H$\alpha$ (top-left
panel), 4000$\AA$ break (D$_{\rm n}(4000\AA)$) (top-right panel), gas-phase
oxygen abundance (12+log(O/H)) (bottom-left panel) and FIR color
($f_\nu(60\micron)/f_\nu(100\micron)$) (bottom-right panel). See Figure
\ref{fuvuvcolorresi.para1.eps} for the explanation of the symbols.}
\label{fuvuvcolorresi.para2.eps} \end{figure}

\begin{figure}
\epsscale{}
\plotone{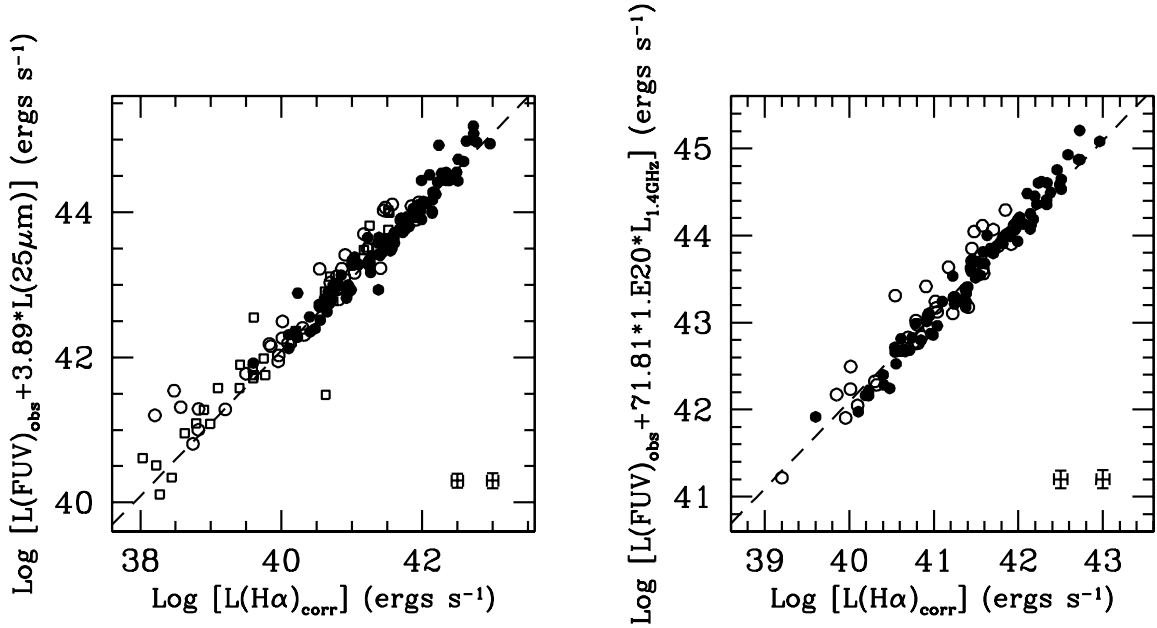}
\caption{Linear combinations of the observed FUV luminosities with 25\um\
luminosities (left panel) and 1.4GHz luminosities (right panel) compared to the
Balmer-corrected H$\alpha$ luminosities, with the scaling coefficient derived
by matching the combined FUV and 25\um\ or 1.4GHz luminosities with the
IRX-corrected FUV luminosities.  The solid circles represent MK06 galaxies, the
open circles denote SINGS galaxies with the integrated measurements while the
open squares denote the central 20\arcsec\ $\times$ 20\arcsec\ regions of SINGS
galaxies. The dashed line denotes the predicted value by the STARBURST99
synthesis model for a constant star formation history, solar metallicity and
Kroupa IMF at age 100 Myr.  The error bars in the bottom-right corner denote
median errors for the MK06 sample (right) and the SINGS sample (left).}
\label{lumifuv25umradio.lha.eps} \end{figure}

\begin{figure}
\epsscale{}
\plotone{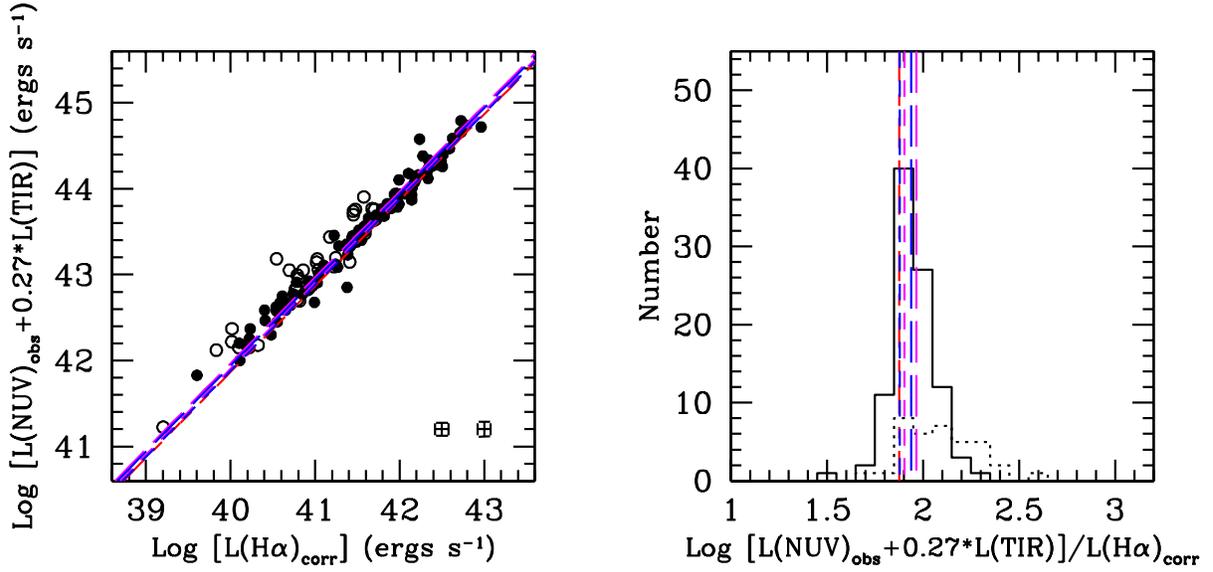}
\caption{{\it left panel:} Combined NUV and TIR luminosities as a function of
Balmer decrement ratio corrected H$\alpha$ luminosities for MK06 sample (solid
circles) and SINGS sample (open circles). The error bars in the 
bottom-right corner denote median errors for the MK06 sample (right) and the
SINGS sample (left). {\it right panel:} Histograms of the
ratios of the TIR/NUV-corrected NUV luminosities to Balmer-corrected H$\alpha$
luminosities for MK06 sample (solid line) and SINGS sample (dotted line) in log
scale.  The color-coded lines represent the same model predictions as those in Figure
\ref{comlumiirx.his.eps}.}
\label{comluminuvirx.lhacor.rat.his.eps} \end{figure}

\begin{figure}
\epsscale{}
\plotone{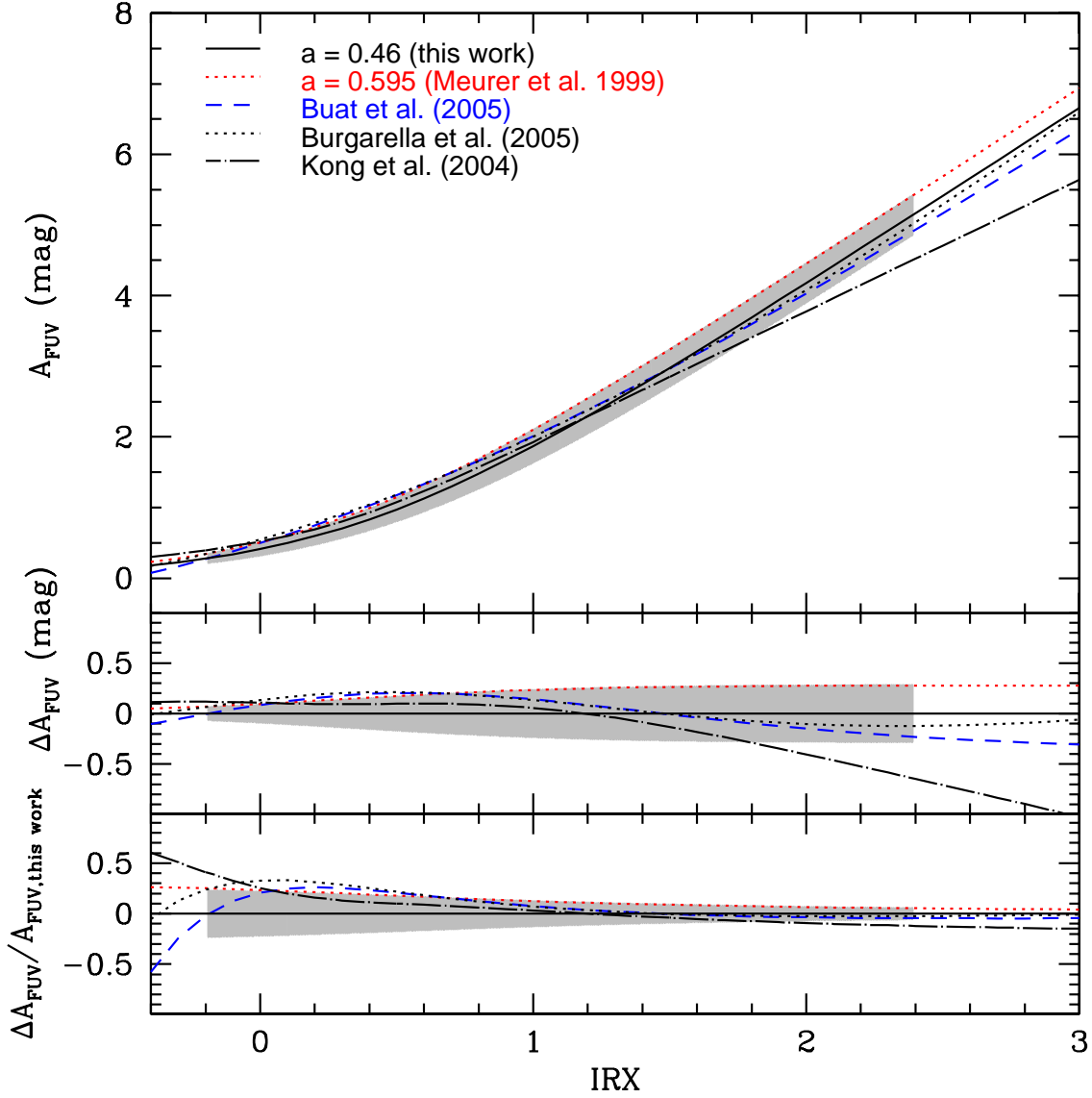}
\caption{Comparison of the A$_{\rm FUV}$ vs. IRX (i.e., $\log {\rm
[(L(TIR)/L(FUV)]}$ relation derived in this paper to those presented by others.
The comparison is made in three different ways: the A$_{\rm FUV}$ itself (top),
the difference in A$_{\rm FUV}$ relative to our calibration (middle) and the
normalized difference (bottom) are plotted as a function of IRX.  The solid
line denotes our result. The shaded region indicates the uncertainty in our
calibration over the range of IRX covered by our sample galaxies. }
\label{compareBuatSBresifuv.eps} \end{figure}

\clearpage

\begin{figure} 
\epsscale{}
\plotone{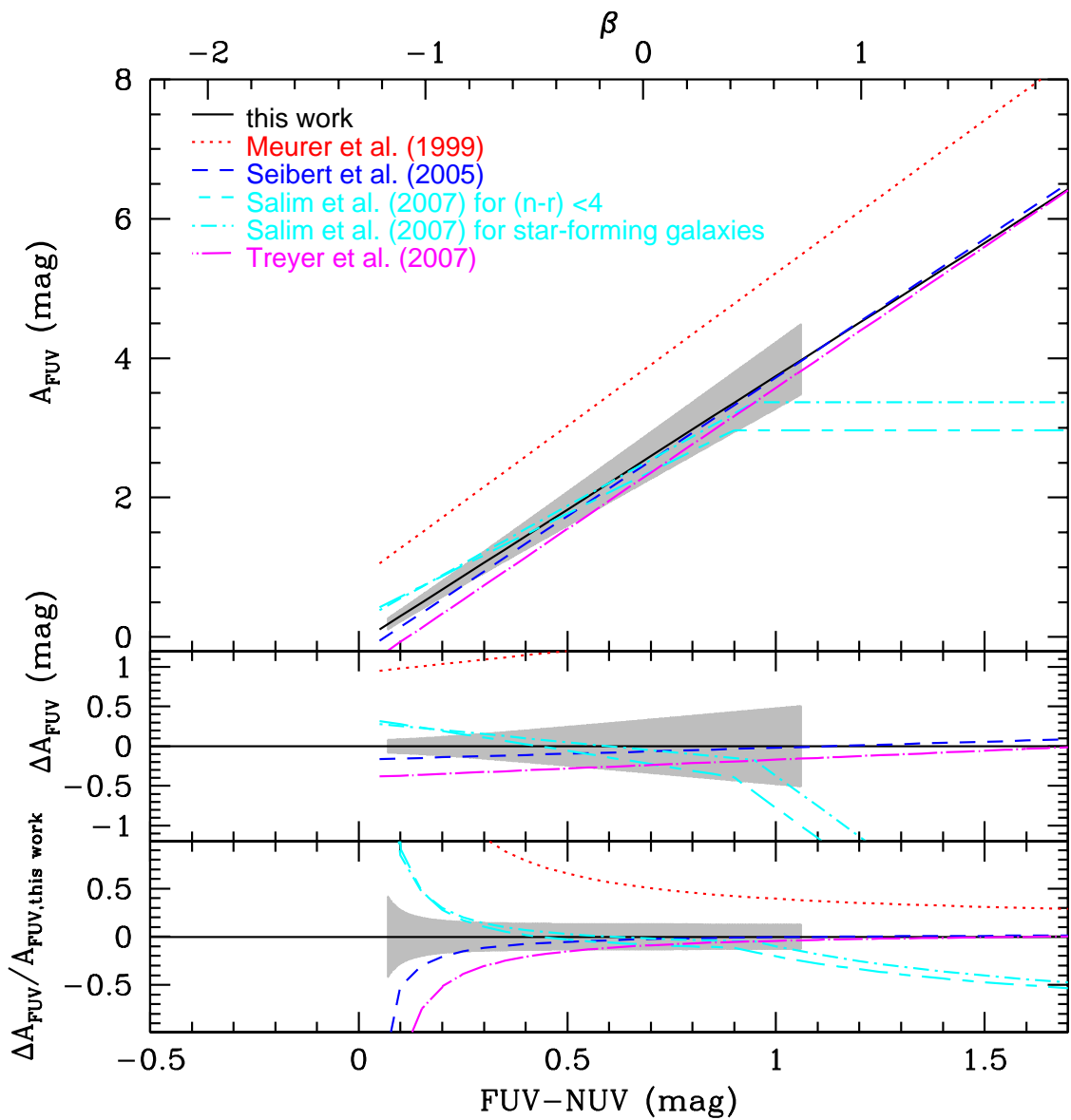}
\caption{Top: Comparison of the A$_{\rm FUV}$ versus FUV-NUV relation obtained in
this work to those presented by others. Middle: The difference in A$_{\rm FUV}$
between the relations given in the literature and that in this paper versus
FUV-NUV. Bottom: The normalized residual versus FUV-NUV.  The solid line denotes
our result. The shaded region denotes the uncertainty in our calibration over
the range of FUV-NUV color spanned by our sample galaxies.}
\label{compareAfuvuvcolorbisecresi.eps} 
\end{figure}

\begin{figure}
\epsscale{}
\plotone{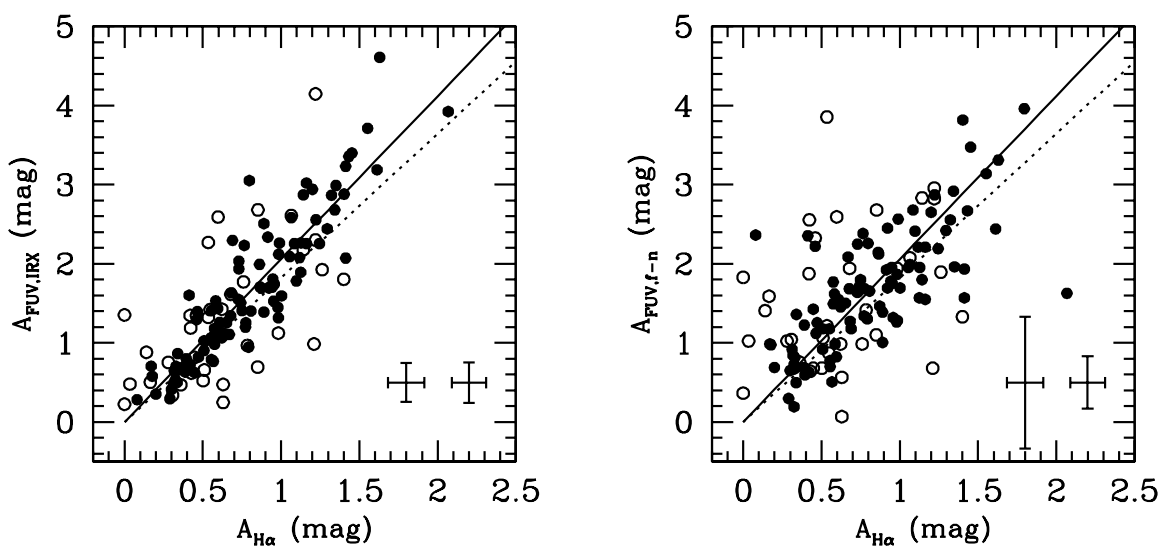}
\caption{IRX (left panel) and FUV-NUV (right panel) derived dust attenuation
in FUV as a function of the dust attenuation in H$\alpha$ line. The solid line
represents the relation defined by our sample galaxies and the dotted line denotes
that given by Calzetti (2001), with the difference in E(B-V)$_\star$ and E(B-V)$_{\rm gas}$
taken into account and adapted to GALEX FUV waveband. The error bars in the 
bottom-right corner denote median errors for the MK06 sample (right) and the
SINGS sample (left).}
\label{Afuv.Aha.Cal.eps}
\end{figure}

\begin{figure}
\epsscale{}
\plotone{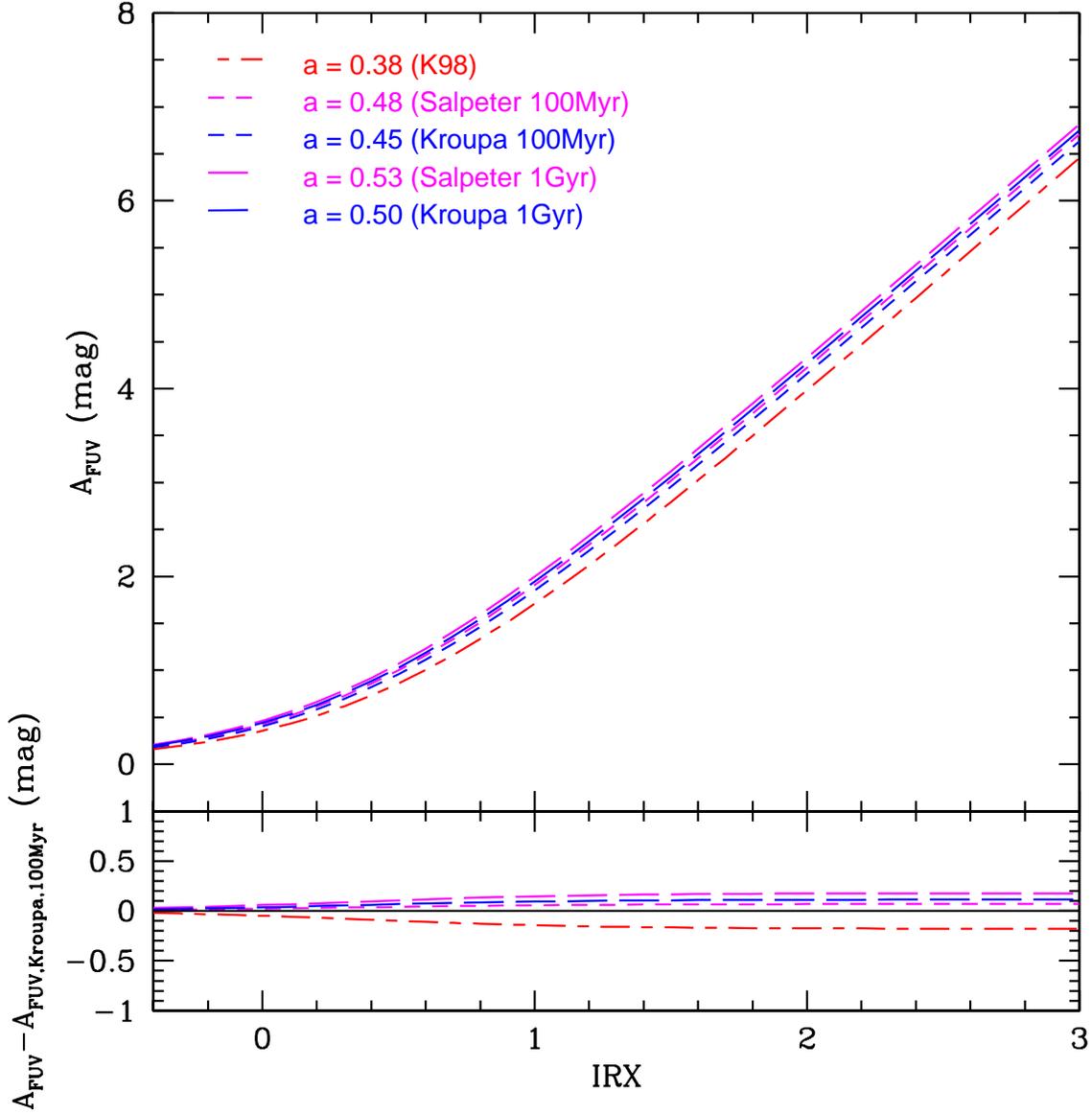}
\caption{Top: A$_{\rm FUV}$ versus IRX (i.e., $\log {\rm [L(TIR)/L(FUV)]}$) relation with
coefficient $a$ derived using the intrinsic FUV to H$\alpha$ luminosity ratio based
on different stellar population assumptions, i.e., IMFs and ages (see text). $a$ is
defined by ${\rm A_{FUV}}=2.5 \log (1 + a_{\rm FUV} \cdot 10^{\rm IRX})$ and is
the approximation of the product of the inverse of the bolometric correction
$\eta_{\rm FUV}$ and the factor ${(1 - {\rm e}^{-{\tau_{\rm FUV}}})} \over {(1 -
{\rm e}^{-\bar{\tau}})}$.
Bottom: The difference in A$_{\rm FUV}$ for different model assumptions.}
\label{compareAfuvirxdiffIMFage.eps}
\end{figure}

\begin{deluxetable}{lrrr}
\tabletypesize{\scriptsize}
\tablecolumns{4}
\tablewidth{0pt}
\tablecaption{GALEX UV photometry for MK06 sample}
\tablehead{
\colhead{Name} & \colhead{Distance}   & \colhead{FUV} & \colhead{NUV} \\
\colhead{} & \colhead{(Mpc)} & \colhead{(mJy)} & \colhead{(mJy)} }
\startdata
ARP256 & 113.9 & 2.83$\pm$0.13\phn & 3.93$\pm$0.11\phn \\
NGC0157 & 24.0 & 13.66$\pm$0.63\phn & 22.41$\pm$0.62\phn \\
NGC0245 & 56.5 & 4.08$\pm$0.23\phn & 6.34$\pm$0.22\phn \\
IC0051 & 24.8 & 1.16$\pm$0.11\phn & 1.99$\pm$0.13\phn \\
NGC0337 & 23.5 & 11.22$\pm$0.52\phn & 19.54$\pm$0.54\phn \\
IC1623 & 84.1 & 6.00$\pm$0.28\phn & 7.81$\pm$0.22\phn \\
MCG-03-04-014 & 139.9 & 0.23$\pm$0.01\phn & 0.52$\pm$0.01\phn \\
NGC0695 & 135.4 & 0.65$\pm$0.03\phn & 1.40$\pm$0.04\phn \\
NGC0877 & 54.6 & 5.11$\pm$0.47\phn & 7.56$\pm$0.40\phn \\
NGC0922 & 43.4 & 14.21$\pm$0.66\phn & 16.39$\pm$0.45\phn \\
NGC0958 & 79.5 & 1.50$\pm$0.07\phn & 2.46$\pm$0.07\phn \\
NGC0959 & 10.9 & 4.52$\pm$0.21\phn & 6.51$\pm$0.18\phn \\
NGC0976 & 59.8 & 1.97$\pm$0.28\phn & 3.20$\pm$0.25\phn \\
NGC1058 & 9.8 & 5.88$\pm$0.42\phn & 9.91$\pm$0.39\phn \\
NGC1084 & 20.0 & 13.77$\pm$0.64\phn & 22.52$\pm$0.62\phn \\
NGC1140 & 21.2 & 9.78$\pm$0.45\phn & 11.76$\pm$0.33\phn \\
NGC1156 & 7.8 & 24.71$\pm$1.14\phn & 30.49$\pm$0.84\phn \\
NGC1359 & 28.0 & 11.54$\pm$0.58\phn & 13.78$\pm$0.42\phn \\
NGC1385 & 21.2 & 17.90$\pm$0.83\phn & 24.76$\pm$0.69\phn \\
IRAS03359+1523 & 147.8 & 0.71$\pm$0.36\phn & 1.25$\pm$0.27\phn \\
NGC1421 & 29.7 & 9.69$\pm$0.66\phn & 14.88$\pm$0.54\phn \\
UGC02982 & 74.5 & 0.63$\pm$2.78\phn & 0.95$\pm$1.90\phn \\
NGC1569 & 2.0 & 372.46$\pm$17.16\phn & 671.21$\pm$18.62\phn \\
NGC2139 & 26.7 & 13.88$\pm$0.72\phn & 20.28$\pm$0.61\phn \\
NGC2337 & 7.9 & 4.84$\pm$0.31\phn & 6.27$\pm$0.25\phn \\
NGC2415 & 57.2 & 5.53$\pm$0.29\phn & 9.41$\pm$0.29\phn \\
NGC2500 & 11.0 & 13.39$\pm$0.62\phn & 15.17$\pm$0.42\phn \\
NGC2537 & 6.9 & 9.07$\pm$0.42\phn & 10.77$\pm$0.30\phn \\
UGC05028 & 56.7 & 1.62$\pm$0.08\phn & 2.45$\pm$0.07\phn \\
NGC2903 & 8.9 & 34.67$\pm$1.60\phn & 60.86$\pm$1.68\phn \\
NGC3239 & 10.5 & 14.91$\pm$0.69\phn & 26.41$\pm$0.73\phn \\
NGC3265 & 25.0 & 0.59$\pm$0.03\phn & 0.97$\pm$0.03\phn \\
UGC05720 & 26.9 & 4.05$\pm$0.19\phn & 5.05$\pm$0.14\phn \\
NGC3344 & 7.9 & 35.41$\pm$1.63\phn & 49.66$\pm$1.37\phn \\
NGC3353 & 19.1 & 4.95$\pm$0.23\phn & 6.10$\pm$0.17\phn \\
NGC3367 & 49.3 & 10.21$\pm$0.52\phn & 15.56$\pm$0.47\phn \\
ARP270 & 30.3 & 17.38$\pm$0.80\phn & 22.15$\pm$0.61\phn \\
NGC3432 & 10.4 & 15.66$\pm$0.75\phn & 22.68$\pm$0.66\phn \\
NGC3442 & 31.6 & 2.91$\pm$0.15\phn & 3.94$\pm$0.13\phn \\
NGC3521 & 8.6 & 17.09$\pm$0.79\phn & 34.79$\pm$0.96\phn \\
NGC3600 & 13.3 & 2.44$\pm$0.11\phn & 3.54$\pm$0.10\phn \\
IC0691 & 23.7 & 1.17$\pm$0.08\phn & 1.58$\pm$0.07\phn \\
NGC3726 & 19.8 & 20.25$\pm$0.99\phn & 28.26$\pm$0.83\phn \\
NGC3738 & 4.9 & 11.49$\pm$0.56\phn & 13.85$\pm$0.40\phn \\
NGC3769 & 19.8 & 3.98$\pm$0.23\phn & 5.89$\pm$0.21\phn \\
UGC06665 & 87.4 & 2.46$\pm$0.11\phn & 3.41$\pm$0.10\phn \\
NGC3870 & 19.8 & 3.33$\pm$0.18\phn & 4.02$\pm$0.13\phn \\
NGC3893 & 15.5 & 13.80$\pm$0.69\phn & 25.50$\pm$0.74\phn \\
NGC3928 & 16.9 & 2.47$\pm$0.15\phn & 3.72$\pm$0.14\phn \\
NGC3949 & 19.8 & 13.60$\pm$0.66\phn & 20.83$\pm$0.60\phn \\
NGC4004 & 55.6 & 2.49$\pm$0.12\phn & 3.61$\pm$0.10\phn \\
ARP244 & 25.0 & 33.06$\pm$1.52\phn & 46.56$\pm$1.29\phn \\
NGC4051 & 15.5 & 15.22$\pm$0.76\phn & 23.94$\pm$0.72\phn \\
NGC4062 & 11.4 & 3.96$\pm$0.20\phn & 7.13$\pm$0.20\phn \\
NGC4085 & 19.8 & 1.40$\pm$0.12\phn & 2.70$\pm$0.13\phn \\
NGC4088 & 19.8 & 11.37$\pm$0.60\phn & 18.43$\pm$0.58\phn \\
NGC4096 & 10.8 & 8.04$\pm$0.37\phn & 12.58$\pm$0.35\phn \\
NGC4100 & 19.8 & 4.24$\pm$0.31\phn & 8.24$\pm$0.32\phn \\
NGC4214 & 2.9 & 52.74$\pm$2.44\phn & 64.17$\pm$1.78\phn \\
NGC4218 & 19.8 & 3.26$\pm$0.16\phn & 4.05$\pm$0.12\phn \\
NGC4254 & 12.9 & 31.13$\pm$1.46\phn & 56.30$\pm$1.57\phn \\
NGC4303 & 10.6 & 43.55$\pm$2.01\phn & 62.09$\pm$1.72\phn \\
NGC4384 & 42.8 & 1.99$\pm$0.13\phn & 3.06$\pm$0.12\phn \\
NGC4414 & 17.7 & 7.46$\pm$0.35\phn & 14.08$\pm$0.39\phn \\
NGC4605 & 5.2 & 21.53$\pm$0.99\phn & 31.48$\pm$0.87\phn \\
NGC4618 & 8.9 & 23.61$\pm$1.09\phn & 30.54$\pm$0.84\phn \\
NGC4625 & 10.2 & 3.88$\pm$0.18\phn & 5.25$\pm$0.14\phn \\
NGC4651 & 20.6 & 9.52$\pm$0.44\phn & 13.98$\pm$0.39\phn \\
NGC4656 & 8.7 & 44.27$\pm$2.04\phn & 48.63$\pm$1.34\phn \\
NGC4666 & 27.6 & 6.14$\pm$0.35\phn & 9.15$\pm$0.34\phn \\
NGC4670 & 23.2 & 7.82$\pm$0.37\phn & 9.38$\pm$0.26\phn \\
NGC4713 & 13.7 & 13.05$\pm$0.61\phn & 17.91$\pm$0.50\phn \\
NGC4900 & 9.0 & 9.86$\pm$0.47\phn & 15.42$\pm$0.43\phn \\
NGC5014 & 23.8 & 0.86$\pm$0.05\phn & 1.55$\pm$0.05\phn \\
MRK0066 & 95.6 & 1.36$\pm$0.09\phn & 1.60$\pm$0.06\phn \\
NGC5194 & 8.0 & 123.02$\pm$5.67\phn & 196.70$\pm$5.43\phn \\
ARP240 & 104.9 & 4.11$\pm$0.25\phn & 6.70$\pm$0.26\phn \\
NGC5253 & 3.1 & 41.14$\pm$1.90\phn & 54.93$\pm$1.52\phn \\
NGC5430 & 49.0 & 2.57$\pm$0.16\phn & 4.05$\pm$0.14\phn \\
NGC5607 & 112.2 & 1.34$\pm$0.11\phn & 2.46$\pm$0.10\phn \\
NGC5591 & 117.6 & 0.68$\pm$0.07\phn & 1.28$\pm$0.08\phn \\
NGC5653 & 58.7 & 1.56$\pm$0.11\phn & 3.02$\pm$0.11\phn \\
NGC5676 & 37.7 & 3.92$\pm$0.31\phn & 8.09$\pm$0.31\phn \\
UGC09560 & 23.7 & 3.30$\pm$0.15\phn & 3.52$\pm$0.10\phn \\
IC1076 & 94.6 & 0.88$\pm$0.10\phn & 1.54$\pm$0.09\phn \\
NGC5996 & 54.6 & 5.63$\pm$0.26\phn & 7.93$\pm$0.22\phn \\
NGC6052 & 74.8 & 4.99$\pm$0.30\phn & 7.60$\pm$0.24\phn \\
NGC6090 & 130.2 & 1.15$\pm$0.05\phn & 1.99$\pm$0.05\phn \\
NGC6926 & 87.6 & 3.08$\pm$0.54\phn & 5.34$\pm$0.45\phn \\
IC5179 & 50.3 & 3.01$\pm$0.18\phn & 5.52$\pm$0.20\phn \\
NGC7316 & 79.0 & 2.58$\pm$0.13\phn & 3.95$\pm$0.11\phn \\
CGCG453-062 & 106.0 & 0.13$\pm$0.02\phn & 0.35$\pm$0.02\phn \\
NGC7624 & 61.1 & 0.80$\pm$0.05\phn & 2.04$\pm$0.06\phn \\
NGC7625 & 24.4 & 1.18$\pm$0.06\phn & 2.78$\pm$0.08\phn \\
NGC7673 & 48.9 & 5.96$\pm$0.28\phn & 7.70$\pm$0.21\phn \\
NGC7678 & 49.7 & 4.87$\pm$0.23\phn & 7.25$\pm$0.20\phn \\
NGC7798 & 34.6 & 3.65$\pm$0.17\phn & 5.59$\pm$0.16\phn \\
\enddata
\end{deluxetable}
\begin{deluxetable}{lccccccc}
\tabletypesize{\scriptsize}
\tablecolumns{8}
\tablewidth{0pc}
\tablecaption{}
\tablehead{\colhead{Model description\tablenotemark{a}} & \colhead{$\log C_{FUV}$\tablenotemark{b}} & \colhead{$\log C_{NUV}$\tablenotemark{b}} & \colhead{$\log C_{H\alpha}$\tablenotemark{b}} & \colhead{$C_{FUV}/C_{H\alpha}$} & \colhead{$C_{NUV}/C_{H\alpha}$} & \colhead{$a_{FUV}$} & \colhead{$a_{NUV}$}}
\startdata
K98 & -43.147 & -42.975 & -41.102 & 0.0090 & 0.0134 & 0.38 & 0.21 \\
Salpeter IMF, ${\rm Z}_\odot$, 100 Myr & -43.170 & -42.959 & -41.056 & 0.0077 & 0.0125 & 0.48 & 0.24 \\
Kroupa IMF, ${\rm Z}_\odot$, 100 Myr & -43.350 & -43.137 & -41.257 & 0.0081 & 0.0132 & 0.45 & 0.22 \\
Salpeter IMF, ${\rm Z}_\odot$, 1 Gyr & -43.207 & -43.023 & -41.056 & 0.0071 & 0.0108 & 0.53 & 0.30 \\
Kroupa IMF, ${\rm Z}_\odot$, 1 Gyr & -43.384 & -43.196 & -41.257 & 0.0075 & 0.0115 & 0.50 & 0.27 \\
\enddata
\tablenotetext{a}{For all cases, a constant star formation history and a mass range of 0.1-100${\rm M}_\odot$ are assumed.}
\tablenotetext{b}{$C_{FUV}$, $C_{NUV}$ and $C_{H\alpha}$ relate to their corresponding SFRs by the relation 
${\rm SFR(\lambda)}=C_\lambda \cdot {\rm
L(\lambda)}$ and are in units of $M_\odot\, yr^{-1}/ergs\, s^{-1}$.}
\end{deluxetable}

\begin{deluxetable}{lcc}
\tabletypesize{\scriptsize}
\tablecolumns{3}
\tablewidth{0pc}
\tablecaption{Summary of Coefficients}
\tablehead{
\colhead{Relation} & \colhead{Coefficient\tablenotemark{b}} &
\colhead{Dispersion}}
\startdata
L(FUV)$_{\rm obs}$ + a*L(TIR) & 0.46$\pm$0.12\phn & 0.09 \\
L(FUV)$_{\rm obs}$ + a*L(25~$\mu$m) & 3.89$\pm$0.15\phn & 0.13  \\
L(FUV)$_{\rm obs}$ + a*L$_{\rm 1.4GHz}$ & 71.81$\pm$5.06 $\times$ 1.E20 & 0.14 \\
L(NUV)$_{\rm obs}$ + a*L(TIR) & 0.27$\pm$0.02\tablenotemark{c} & 0.10 \\
L(NUV)$_{\rm obs}$ + a*L(25~$\mu$m) & 2.26$\pm$0.09\tablenotemark{c} & 0.13  \\
L(NUV)$_{\rm obs}$ + a*L$_{\rm 1.4GHz}$ & 41.75$\pm$2.97\tablenotemark{c} $\times$ 1.E20 & 0.14 \\
\enddata
\tablenotetext{b}{The coefficients $a$ in the combinations of UV and TIR or 25~$\mu$m luminosities are unitless,
and the luminosities in these combinations are in units of $ergs\, s^{-1}$.
The coefficients $a$ in the combinations of UV and 1.4 GHz radio luminosities are in units of $10^{-7}\,Hz$ ,
the UV luminosities are in units of $ergs\, s^{-1}$ and the 1.4 GHz radio luminosities are in units of $w\,Hz^{-1}$. }
\tablenotetext{c}{The error does not include the uncertainties in the reference luminosities, so it is under-estimated.}
\end{deluxetable}


\begin{thebibliography}

\bibitem[]{} Balogh, M. L., Morris, S. L., Yee, H. K. C., Carlberg, R. G., \& Ellingson, E. 1999, \apj, 527, 54

\bibitem[]{} Bell, E. F. 2002, \apj, 577, 150

\bibitem[]{} Bell, E.F. 2003, \apj, 586, 794

\bibitem[]{} Bell, E. F., \& Kennicutt, R. C., Jr. 2001, \apj, 548, 681

\bibitem[]{} Boissier, S., et al. 2007, \apjs, 173, 524

\bibitem[]{} Boquien, M., et al. 2009, \apj, 706, 553

\bibitem[]{} Brinchmann, J., Charlot, S., White, S. D. M., Tremonti, C., Kauffmann, G.,
       Heckman, T., \& Brinkmann, J. 2004, \mnras, 351, 1151

\bibitem[]{} Bruzual, G. 1983, \apj, 273, 106

\bibitem[]{} Bruzual, G., \& Charlot, S. 2003, \mnras, 344, 1000

\bibitem[]{} Buat, V., Boselli, A., Gavazzi, G., \& Bonfanti, C. 2002, \aap, 383, 801

\bibitem[]{} Buat, V., Donas, J., Milliard, B., \& Xu, C. 1999, \aap, 352, 371

\bibitem[]{} Buat, V. et al. 2010, \mnras, 409, 1

\bibitem[]{} Buat, V. et al. 2005, \apj, 619, L51

\bibitem[]{} Buat, V., \& Xu, C. 1996, \aap, 306, 61

\bibitem[]{} Burgarella, D., Buat, V., \& Iglesias-P\'aramo, J. 2005, \mnras, 360, 1413

\bibitem[]{} Calzetti, D. 1997, \aj, 113, 162

\bibitem[]{} Calzetti, D. 2001, \pasp, 113, 1449

\bibitem[]{} Calzetti, D., Armus, L., Bohlin, R. C., Kinney, A. L., Koornneef, J., \& Storchi-Bergmann, T. 2000, \apj, 533, 682

\bibitem[]{} Calzetti, D., et al. 2005, \apj, 633, 871

\bibitem[]{} Calzetti, D., et al. 2010, \apj, 714, 1256

\bibitem[]{} Calzetti, D., Kinney, A. L., \& Storchi-Bergmann, T. 1994, \apj, 429, 582

\bibitem[]{} Cardelli, J. A., Clayton, G. C., \& Mathis, J. S. 1989, \apj, 345, 245

\bibitem[]{} Charlot, S., \& Fall, S. M. 2000, \apj, 539, 718

\bibitem[]{} Cortese, L., et al. 2006, \apj, 637, 242

\bibitem[]{} Cortese, L., Boselli, A., Franzetti, P., Decarli, R.,
       Gavazzi, G., Boissier, S., \& Buat, V. 2008, \mnras, 386, 1157

\bibitem[]{} Dale, D. A., \& Helou, G. 2002, ApJ, 576, 159

\bibitem[]{} Dale, D.A., et al. 2007, \apj, 655, 863

\bibitem[]{} Dale, D.A., et al. 2009, \apj, 703, 517

\bibitem[]{} Feigelson, E. D., \& Babu, G. J. 1992, \apj, 397, 55

\bibitem[]{} Gil de Paz, A., et al. 2007, \apjs, 173, 185

\bibitem[]{} Gordon, K.D., Clayton, G.C., Witt, A.N., \& Misselt, K.A.
       2000, \apj, 533, 236

\bibitem[]{} Granato, G. L., Lacey, C. G., Silva, L., Bressan, A., Baugh, C. M., Cole, S., \& Frenk, C. S. 2000, \apj, 542, 710

\bibitem[]{} Goldader, J. D., Meurer, G., Heckman, T. M., Seibert, M., Sanders, D. B., Calzetti, D., \& Steidel, C. C. 2002, \apj, 568, 651

\bibitem[]{} Heckman, T. M., Robert, C., Leitherer, C., Garnett, D. R., \& van der Rydt, F. 1998, \apj, 503, 646

\bibitem[]{} Howell, J. H., et al. 2010, \apj, 715, 572

\bibitem[]{} Iglesias-P\'aramo, J., et al. 2006, ApJS, 164, 38

\bibitem[]{} Isobe, T., Feigelson, E. D., Akritas, M. G., \& Babu, G. J. 1990, \apj, 364, 104

\bibitem[]{} Johnson, B.D., et al. 2007a, \apjs, 173, 377

\bibitem[]{} Johnson, B.D., et al. 2007b, \apjs, 173, 392

\bibitem[]{} Kauffmann, G., et al. 2003, \mnras, 341, 33

\bibitem[]{} Kennicutt, R.C. 1998, \araa, 36, 189

\bibitem[]{} Kennicutt, R. C., Tamblyn, P., \& Congdon, C. W. 1994, \apj, 435, 22

\bibitem[]{} Kennicutt, R.C., et al. 2003, \pasp, 115, 98

\bibitem[]{} Kennicutt, R.C., et al. 2009, \apj, 703, 1672

\bibitem[]{} Kewley, L. J., Heisler, C. A., Dopita, M. A., \& Lumsden, S. 2001, \apjs, 132, 37

\bibitem[]{} Kinney, A.L., Bohlin, R. C., Calzetti, D., Panagia, N., \&
       Wyse, R. F. G. 1993,\apjs, 86, 5

\bibitem[]{} Kong, X., Charlot, S., Brinchmann, J., \& Fall, S.M. 2004, \mnras, 349, 769

\bibitem[]{} Kroupa, P., \& Weidner, C. 2003, \apj, 598, 1076

\bibitem[]{} Lee, J. C., et al. 2009, \apj, 706, 599

\bibitem[]{} Leitherer, C., \& Heckman, T. M. 1995, \apjs, 96, 9

\bibitem[]{} Leitherer, C., et al. 1999, \apjs, 123, 3

\bibitem[]{} Li, A., \& Draine, B. T. 2001, \apj, 554, 778

\bibitem[]{} Martin, D. C., et al. 2005, \apjl, 619, L1

\bibitem[]{} Madau, P., Pozzetti, L., \& Dickinson, M. 1998, \apj, 498, 106

\bibitem[]{} Meurer, G. R., Heckman, T. M., \& Calzetti, D. 1999, \apj, 521, 64

\bibitem[]{} Meurer, G. R., Heckman, T. M., Leitherer, C., Kinney, A., Robert, C., \& Garnett, D. R. 1995, \aj, 110, 2665

\bibitem[]{} Meurer, G. R., et al. 2009, \apj, 695, 765

\bibitem[]{} Morrissey, P., et al. 2007, \apjs, 173, 682

\bibitem[]{} Moustakas, J., \& Kennicutt, R.C. 2006, \apjs, 164, 81 (MK06)

\bibitem[]{} Moustakas, J., Kennicutt, R.C., \& Tremonti, C.A. 2006, \apj, 642, 775

\bibitem[]{} O'Donnell, J.E. 1994, \apj, 422, 158

\bibitem[]{} Panuzzo, P., Granato, G. L., Buat, V., Inoue, A. K., Silva, L., Iglesias-P\'aramo, J., \& Bressan, A. 2007, \mnras, 375, 640

\bibitem[]{} Salim, S., et al. 2009, \apj, 700, 161

\bibitem[]{} Salim, S., et al. 2007, \apjs, 173, 267

\bibitem[]{} Salpeter, E.E. 1955, \apj, 121, 161

\bibitem[]{} Sauvage, M., \& Thuan, T.X. 1992, \apj, 396, 69

\bibitem[]{} Schlegel, D. J., Finkbeiner, D. P., \& Davis, M. 1998, \apj, 500, 525

\bibitem[]{} Seibert, M., et al. 2005, \apj, 619, 55

\bibitem[]{} Takeuchi, T. T., Buat, V., Heinis, S., Giovannoli, E., Yuan, F.-T., Iglesias-P\'aramo, J., Murata, K. L., \& Burgarella, D. 2010, \aap, 514, 4

\bibitem[]{} Treyer, M., et al. 2007, \apjs, 173, 256

\bibitem[]{} V\'azquez, G. A., \& Leitherer, C. 2005, \apj, 621, 695

\bibitem[]{} Wang, B., \& Heckman, T. M. 1996, \apj, 457, 645

\bibitem[]{} Witt, A. N., \& Gordon, K. D. 2000, \apj, 528, 799

\bibitem[]{} Witt, A. N., Thronson, H. A., Jr., \& Capuano, J. M., Jr. 1992, \apj, 393, 611

\bibitem[]{} Zhu, Y. N., Wu, H., Cao, C., \& Li, H. N. 2008, \apj, 686, 155

\end{thebibliography}
\end{document}